\documentclass[journal=jpcbfk,manuscript=article,layout=twocolumn]{achemso}

\usepackage{chemformula} % Formula subscripts using \ch{}
\usepackage[T1]{fontenc} % Use modern font encodings
\usepackage{graphicx}
\usepackage{amsmath}
\usepackage{amssymb}

\author{Yuichi Togashi}
\email{togashi@hiroshima-u.ac.jp}
\affiliation[Hiroshima University]
{Research Center for the Mathematics on Chromatin Live Dynamics (RcMcD); Department of Mathematical and Life Sciences, Graduate School of Science, Hiroshima University, 1-3-1 Kagamiyama, Higashi-Hiroshima, Hiroshima 739-8526, Japan}
\alsoaffiliation[RIKEN BDR]
{RIKEN Center for Biosystems Dynamics Research (BDR), 3-10-23 Kagamiyama, Higashi-Hiroshima, Hiroshima 739-0046, Japan}
\alsoaffiliation[Osaka University]
{Cybermedia Center, Osaka University, 5-1 Mihogaoka, Ibaraki, Osaka 567-0047, Japan}
\altaffiliation{ORC ID: 0000-0003-0032-2424}
\phone{+81 82 424 7373}
\fax{+81 82 424 7373}

\title[Modeling of Nano-/Micro-machine Crowds]
  {Modeling of Nano-/Micro-machine Crowds: Interplay between the Internal State and Surroundings}


\SectionNumbersOn

\begin{document}

\onecolumn

\begin{abstract}
The activity of biological cells is primarily based on chemical reactions and
typically modeled as a reaction-diffusion system.
Cells are, however, highly crowded with macromolecules, including
a variety of molecular machines such as enzymes.
The working cycles of these machines are often coupled with their
internal motion (conformational changes).
In the crowded environment of a cell, motion interference
between neighboring molecules is not negligible,
and this interference can affect the reaction dynamics through machine operation.
To simulate such a situation,
we propose a reaction-diffusion model consisting of particles
whose shape depends on an internal state variable,
for crowds of nano- to micro-machines.
The interference between nearby particles is naturally introduced
through excluded volume repulsion.
In the simulations,
we observed segregation and flow-like patterns enhanced by crowding out
of relevant molecules, as well as molecular synchronization waves and
phase transitions.
The presented model is simple and extensible for diverse molecular machinery,
and may serve as a framework to study the interplay between
the mechanical stress/strain network and the chemical reaction network
in the cell.
Applications to more macroscopic systems, e.g., crowds of cells,
are also discussed.
\end{abstract}

\section{Introduction}

Biological cells maintain their activity by utilizing a huge variety of
macromolecules, such as nucleic acids and proteins.
Proteins working as molecular machines are of particular importance,
because they function as, for instance, motors, transporters, or enzymes,
to facilitate active processes.
Since these processes are mostly supported by chemical reactions,
they are often modeled as reaction-diffusion systems. Thus, they are
typically represented by reaction-diffusion equations,
i.e., partial differential equations of concentrations.

However, the operation of molecular machines is coupled with
their conformational changes, which are generally slow,
taking microseconds to seconds to finish.
Hence, the dynamics or delay in operation cannot always be ignored.
Moreover, the number of each species of protein molecules is
not always large, in which case we must consider discrete molecules.
Considering these aspects,
in our previous works on reaction-diffusion systems with enzymes,
we used a phase model for each enzyme to study the effects of
the delay in the reaction cycle, which is related to conformational
changes of the enzyme \cite{Casagrande2007,Togashi2015} (the reaction
scheme was originally proposed in \cite{Stange1998}).
A variety of spatiotemporal patterns, such as traveling waves
and spirals, were observed, and the effects of fluctuations on
pattern formation were also investigated.

Nevertheless, there are properties of enzymes and the intracellular environment
that were not yet implemented in our previous model.
Although discrete molecules in the space were considered,
their shapes and excluded volume were ignored.
The environment in the cell is crowded with macromolecules,
including molecular machines.
Indeed, crowding effects in the cell have been 
studied extensively by others (for reviews, see, e.g., \cite{Elcock2010,Mourao2014,Rivas2016}).
However, molecular machines exert internal motion for the operation.
If they work in a crowded environment, the motion can affect, and be affected by, the surroundings.
This means that the operation of molecular machines can interfere
with each other through the environment, which may in turn change
the collective reaction dynamics in the machinery.

Although there are many studies that pertain to the crowding effects
on diffusion or structural stability of macromolecules,
the interplay between machines and their surroundings have not been
well considered.
Here we propose a simple model to consider the interplay between
the internal states of molecular machines and the environment,
by considering the shapes of the machines. We
further demonstrate the effects on spatiotemporal pattern formation in the reaction-diffusion process.
This model is easily extensible and applicable to various
mechanochemical behavior in crowded molecular machinery
such as chromatin structures in the cell nucleus.
Finally, we discuss applications for understanding the collective behavior of more macroscopic systems,
such as crowds of cells or artificial micro-machines.

\section{Methods}

\begin{figure}
\begin{center}
\includegraphics[width=82.5mm]{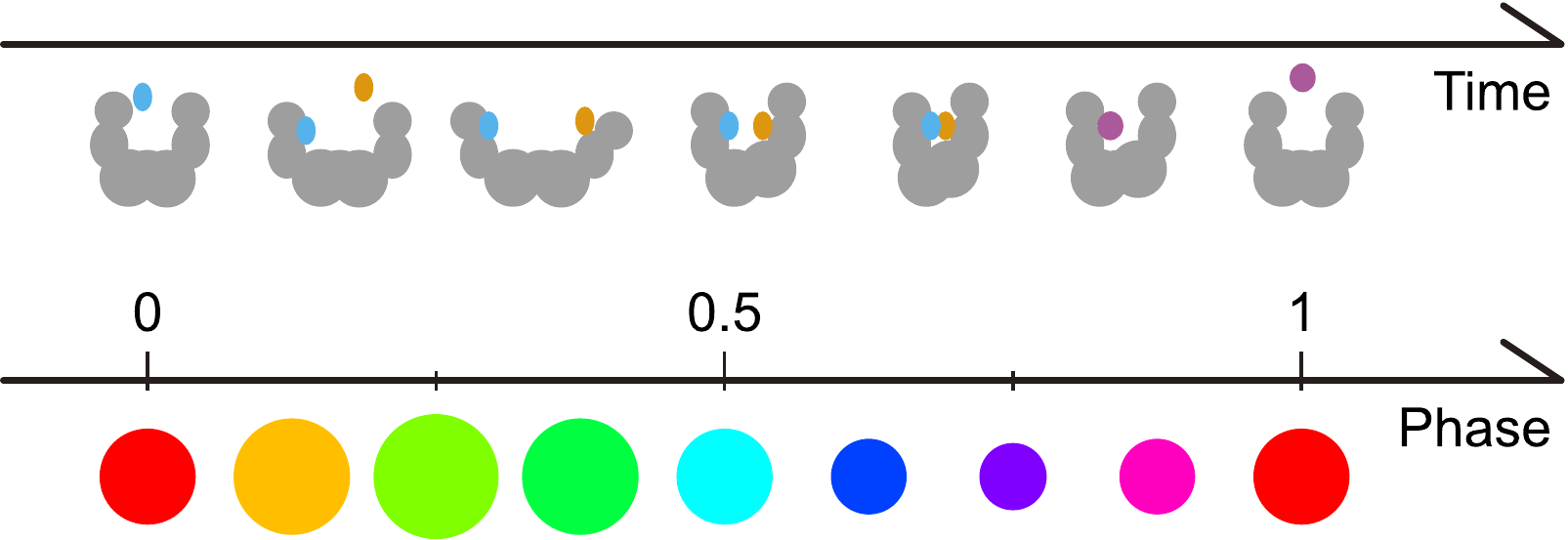}
\end{center}
\caption{Concept of the machine model.
To consider conformational changes coupled with the machine operation (top),
each machine is modeled as a particle (circular disk in 2D)
whose radius changes along the operation cycle (bottom).
For simplicity, we assume that the state of the machine is represented
by a single phase variable,
and the radius is determined as a function of the phase.
When the operation cycle starts, the phase is reset to $0$,
and then proceeds.
When the phase reaches $1$, the machine returns to a resting state.
}
\label{fig-model1}
\end{figure}

We constructed a model consisting of particles
(i.e., circular disks in two-dimensional space).
Both the machines (e.g., enzymes) and other objects (e.g., substrates and products)
are modeled as particles diffusing in the space.
Each machine has an internal state represented by a phase variable,
on which the shape (radius) of the particle depends (Fig. \ref{fig-model1}).
The interaction between neighboring particles is introduced as
repulsive forces through the exclusion volume.
Particles with oscillating radii were previously adopted
by Awazu \cite{Awazu2007} and Tjhung and Berthier \cite{Tjhung2017}.
Our model is similar to their models, except in the following ways:
(i) the machine stops (i.e., returns to a resting state) after each cycle of operation, and 
(ii) the progress of the phase is affected by the surroundings.

With the exception of (i), the proposed model is analogous to studies of mobile oscillators,
particularly those in which interactions between the oscillators
and their consequent motion are
affected (directly or indirectly) by the phases of the oscillators
\cite{Sawai1998,Zanette2004,Tanaka2007,Iwasa2010,Iwasa2011,Iwasa2012,Prignano2013,Iwasa2017,OKeeffe2017}
(see also \cite{Banerjee2017};
similar systems were constructed with dynamical systems other than oscillators,
such as maps \cite{Shibata2003};
other works assumed random motion independent of the oscillator phases,
e.g. \cite{Uriu2010,Fujiwara2011,Majhi2017}).

\subsection{Common settings}
\label{sect-common}

We consider the long-time behavior of the system,
and assume that the motion of the particles is in the overdamped limit
(i.e., the velocity is proportional to the force).
The position (center coordinate) $\mathbf{x}_{i}$ of particle $i$ is
represented by the overdamped Langevin equation:
\begin{equation}
\frac{dx_{i}}{dt} = - \mu_{i} \frac{\partial U}{\partial x_{i}} + \eta_{i}(t),
\label{eqn-langevin-x}
\end{equation}
where $U$ is the potential energy of the interactions between the particles,
and $\eta_{i}(t)$ is white noise with
$\left\langle \eta_{i}(t) \right\rangle = 0$,
$\left\langle \eta_{i}(t) \eta_{j}(t^{\prime}) \right\rangle = 2 \mu_{i} \sigma \delta_{ij} \delta(t-t^{\prime})$.
Parameter $\mu_{i}$ is the mobility of the particle, and
$\sigma$ specifies the intensity of noise (corresponding to $k_\mathrm{B} T$ for thermal fluctuations).

We assume that each particle is a semi-hard circular disk in two-dimensional space,
and introduce a repulsive 12-6 Lennard-Jones potential $u_{ij}$ between nearby particles,
such that
\begin{equation}
u_{ij} = H(2^{\frac{1}{6}} R_{ij} - d_{ij}) \left[ \left( \frac{R_{ij}}{d_{ij}} \right)^{12} - \left( \frac{R_{ij}}{d_{ij}} \right)^{6} + \frac{1}{4} \right],
\label{eqn-potential-1}
\end{equation}
where $d_{ij} = \left| \mathbf{x}_{i} - \mathbf{x}_{j} \right|$, and $R_{ij} = r_{i} + r_{j}$;
$r_{i}$ is the radius of particle $i$, and $H(x)$ is the unit step function $H(x \ge 0) = 1$.
The repulsive force acts only when the center-to-center distance $d_{ij}$ between
two particles is less than $2^{\frac{1}{6}}$ times the sum $R_{ij}$ of their radii.
Below, we use the term ``in contact'' when the repulsive force is effective,
i.e., $d_{ij} < 2^{\frac{1}{6}} R_{ij}$ (\underline{not} $d_{ij} < R_{ij}$).
The total potential energy is
\begin{equation}
U = \sum_{i} \sum_{j \ne i} u_{ij}.
% H(2^{\frac{1}{6}} R_{ij} - d_{ij}) \left[ \left( \frac{R_{ij}}{d_{ij}} \right)^{12} - \left( \frac{R_{ij}}{d_{ij}} \right)^{6} + \frac{1}{4} \right].
\label{eqn-potential}
\end{equation}

We now model the operation of each machine.
Initially, the machines are in a resting (free) state.
The operation cycle is initiated stochastically 
(where the conditions for the initiation depend on the model, as described in Sections \ref{sect-model1} and \ref{sect-model2}).
Each machine particle $i$ in the operation cycle has an internal state
represented by a single phase variable $\phi_{i}$ ($0 \le \phi_{i} \le 1$).
and its radius $r_{i}$ changes as a function of $\phi_{i}$.
When the operation cycle starts, $\phi_{i}$ is reset to $0$, and then proceeds.
When $\phi_{i}$ reaches $1$, the machine returns to a resting state, ready for the next cycle.
During the operation, the size of the particle changes,
representing the structural changes of the machine.
For simplicity, we assume a simple sine curve for the radius:
\begin{equation}
r_{i} = r_{0} \left( 1 + A \sin 2\pi \phi_{i} \right);
\label{eqn-radius}
\end{equation}
$A$ ($0 \le A < 1$) represents the extent of deformation.
For machines in a resting state, the radius $r_{i} = r_{0}$.

In the same way as our previous works on enzymes \cite{Casagrande2007,Togashi2015},
we consider a stochastic diffusional drift process (biased Brownian motion)
for state $\phi_{i}$, such that
\begin{equation}
\frac{d \phi_{i}}{dt} = v + \eta'_{i}(t),
\label{eqn-langevin-phi-0}
\end{equation}
where $v$ is the mean velocity, and $\eta'_{i}(t)$ is white noise with
$\left\langle \eta'_{i}(t) \right\rangle = 0$,
$\left\langle \eta'_{i}(t) \eta'_{j}(t^{\prime}) \right\rangle = 2 \mu_{\phi} \sigma_{\phi} \delta_{ij} \delta(t-t^{\prime})$.
Here, two parameters are introduced:
$\mu_{\phi}$ is the ``mobility'', or susceptibility, of the phase, and
$\sigma_{\phi}$ specifies the intensity of internal noise affecting the operation.
(In our previous model \cite{Togashi2015}, a single parameter $\sigma = \mu_{\phi} \sigma_{\phi}$ was
used for the intensity of intramolecular fluctuations.)

In the current case, the progress of the machine operation is
also affected by the surroundings.
For example, suppose that the radius $r_{i}$ is increasing with $\phi_{i}$ (i.e.,
the machine conformation is opening).
If the machine is pressed from the neighbors, then its conformational change
will be hindered, and its operation will be deterred.
In this model, this effect is naturally introduced through the potential $U$ and
the susceptibility $\mu_{\phi}$.
Here, we introduce into Eq. (\ref{eqn-langevin-phi-0})  a virtual ``force'' from
the potential acting on the phase.
In the same way as Eq. (\ref{eqn-langevin-x}),
\begin{equation}
\frac{d \phi_{i}}{dt} = v - \mu_{\phi} \frac{\partial U}{\partial \phi_{i}} + \eta'_{i}(t).
\label{eqn-langevin-phi}
\end{equation}
As the potential $U$ (eqs. (\ref{eqn-potential-1}) and (\ref{eqn-potential})) depends on the radii $r_{i}$,
\begin{equation}
\frac{\partial U}{\partial \phi_{i}} = \frac{\partial U}{\partial r_{i}} \cdot \frac{\partial r_{i}}{\partial \phi_{i}}
\label{eqn-phase-force}
\end{equation}
(Note: This ``force'' may also correspond to realistic forces
affecting the internal deformation of the machine.
The constant drift $v$ is unaffected by $\mu_{\phi}$. Hence, smaller $\mu_{\phi}$ may
imply stiffer and more powerful machines.)
Substituting the derivatives of Eqs. (\ref{eqn-potential}) and (\ref{eqn-radius})
to Eq. (\ref{eqn-phase-force}), we obtain the explicit form:
\begin{equation}
\frac{\partial U}{\partial \phi_{i}} = 2\pi r_{0} A \cos 2\pi \phi_{i}
\sum_{j \ne i} H(2^{\frac{1}{6}} R_{ij} - d_{ij}) \left[ \frac{12}{R_{ij}} \left( \frac{R_{ij}}{d_{ij}} \right)^{12} - \frac{6}{R_{ij}} \left( \frac{R_{ij}}{d_{ij}} \right)^{6} \right].
\label{eqn-phase-force-1}
\end{equation}
Note that it is nonzero only when particle $i$ is in contact with another particle.

We fix $v = 1$ throughout, such that the unit-time corresponds
to the typical time required for the operation cycle.
(We can rescale the time without loss of generality.)
For particles other than machines, the radius $r_{i}$ is fixed for each type of particle.
We assume $r_{0} = 0.5$, $\mu_{i} = 1$ for the machines,
and $r_{i} = 0.1$, $\mu_{i} = 5$ for the other particles.
The periodic boundary condition is used.
The turnover rate is then calculated as the number of operation cycles finishes
in the designated timespan, per unit-time and machine.

The radial distribution function $\rho (d)$ is obtained as the average density
at the distance range between $d - \frac{\Delta d}{2}$ and $d + \frac{\Delta d}{2}$ (in the current 2D case,
the number of particles is divided by $2 \pi d \Delta d$), normalized by the total density.
To characterize local synchronization, neglecting the resting state,
phase correlation $\chi (d)$ is calculated as
cosine similarity $\cos \left( \phi_{i} - \phi_{j} \right)$ averaged over
all the pairs $(i,j)$ of machines in the cycle at the distance range between $d - \frac{\Delta d}{2}$ and $d + \frac{\Delta d}{2}$;
$-1 \le \chi (d) \le 1$.
$\chi (d) = 1$ if $\phi_{i}$ of these machines are uniform (completely synchronized),
and $\chi (d) = 0$ if random.
Snapshots at intervals 0.1 for $50 \le t \le 100$ and 20 trials (10020 snapshots in total)
were used for the calculation of $\rho (d)$ and $\chi (d)$. $\Delta d = 0.01$.

\subsection{Model 1: Basic model}
\label{sect-model1}

First, we consider systems consisting exclusively of machine particles.
Each particle in the resting state starts the operation cycle
independently and randomly at rate $\alpha$ (i.e., the waiting time
distributes exponentially).
Parameters $\alpha=1$, $\mu = 1$, $\mu_{\phi} = 1/(2\pi)$,
and simulation timestep $\Delta t = 1 \times 10^{-4}$ were used.

%ver. 4.31a
For the dependence of the turnover rate on $\sigma_{\phi}$ (Fig. \ref{fig-turnover-sigmaphi}) and $\sigma$ (Fig. \ref{fig-turnover-sigma}),
the system size $L_{x} = L_{y} = 64$ and the number of the machine particles $N = 4096$
were used.
After waiting for initial relaxation, the turnover rate was measured for $400 < t \le 500$.
%ver. 4.32a
For the dependence of the turnover rate on the density (Fig. \ref{fig-turnover-density})---i.e., to prepare a system at high density--- we initially set $L_{x} = L_{y} = 96$, randomly distributed $N = 4096$ machine particles
with the minimum center-to-center distance $d_{initmin} = 1$, and started the simulation.
Then, for $0 < t < 100$, $L_{x}$ and $L_{y}$ were exponentially decreased to the target value,
linearly rescaling the particle coordinates (like the Berendsen barostat).
The turnover rate was averaged over $150 < t \le 300$, and
$\sigma = 1$ and  $\sigma_{\phi} = 0$ were used.
In both cases, $A=0.5$, and
the mean and standard deviation over 10 trials are shown.

\subsubsection{Additional couplings of the phases}
\label{sect-phasecoupling}

If we use this model for more macroscopic objects than molecular machines,
such as cells or artificial micromachines,
there may be other kinds of communication or interference between
the machines.
These can be introduced as additional couplings between the phase variables.
In the example in Fig. \ref{fig-snapshot-kuramoto}, as a demonstration,
we assumed Kuramoto-type couplings between the phase variables \cite{Kuramoto1984,Acebron2005},
and Eq. (\ref{eqn-langevin-phi}) was modified as
\begin{equation}
\frac{d \phi_{i}}{dt} = v - \frac{\partial U}{\partial \phi_{i}} + \epsilon \sum_{j \ne i} u_{ij} \sin 2\pi \left( \phi_{j} - \phi_{i} \right)  + \eta'_{i}(t).
\label{eqn-kuramoto}
\end{equation}
Unlike Kuramoto's original coupled phase oscillators,
the coupling acts only when both particles are in the operation cycle (i.e., when not in a resting state)
and when they are in contact. The coupling strength depends on their distance.
This can be regarded as an oscillator network with dynamic coupling strengths
depending on the geometry \cite{Rodrigues2016}.
Furthermore, $L_{x} = L_{y} = 64$, $N = 4096$,
$A=0.3$, $\sigma = 0.04$, $\sigma_{\phi}=0.04 /(2\pi)$, and $\epsilon = 25/(2\pi)$ were used.

\subsection{Model 2: Explicit substrate model}
\label{sect-model2}

\begin{figure}
\begin{center}
\includegraphics[width=82.5mm]{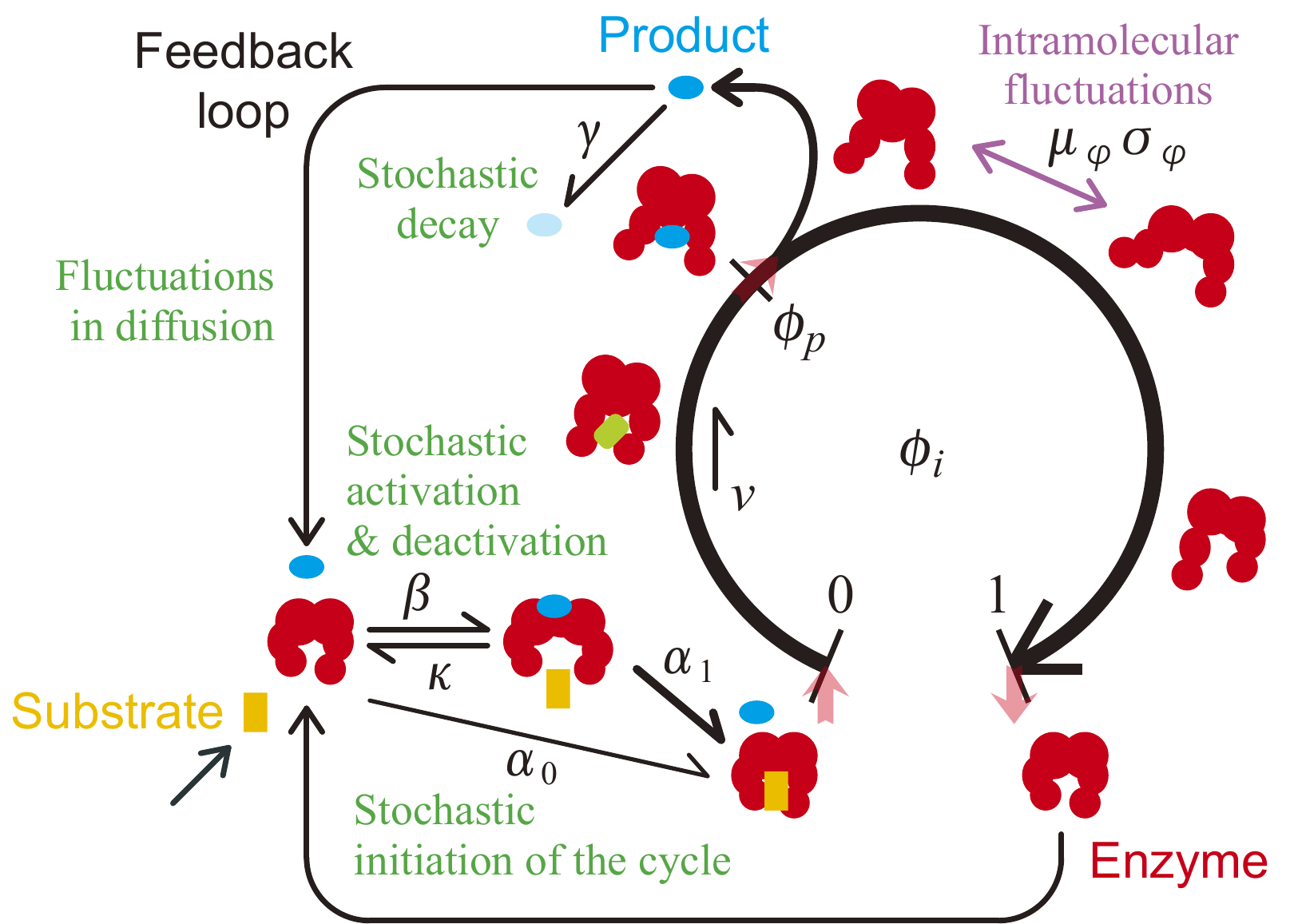}
\end{center}
\caption{Schematic diagram of the reactions in Model 2.
The state of each machine $i$ is represented by a phase variable $\phi_{i}$.
Adapted from \cite{Togashi2015}.
}
\label{fig-model2}
\end{figure}

\begin{figure}
\includegraphics[width=82.5mm]{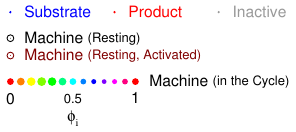}
\caption{Types of particles.
This legend is common to Figs. \ref{fig-snapshot-kuramoto},
\ref{fig-snapshot-waves}, \ref{fig-snapshot-paths}, and \ref{fig-snapshot-inactive}.
Colors of machine particles indicate phase $\phi_{i}$ in the operation cycle.
}
\label{fig-snapshot-legend}
\end{figure}

Next, we combined the model with the enzymatic reaction model
presented in our previous work \cite{Togashi2015} (Fig. \ref{fig-model2}).
% We also included the substrate and product of the enzyme.
When a substrate (S) binds to the enzyme (machine; E), the enzyme stochastically
starts the operation cycle,
in which phase $\phi_{i}$ proceeds as introduced in Eq. (\ref{eqn-langevin-phi}).
Additionally, when $\phi_{i}$ reaches $\phi_{p}$, a product (P) is released.
The product can allosterically bind to and unbind from an enzyme in a resting state,
which alters the cycle initiation rate $\alpha$ (i.e., the regulatory product).

In our previous study, enzymes were fixed on the surface,
and the product could diffuse.
Substrates were not explicitly considered, and their concentration was assumed to be constant.
With the proposed model, the enzymes (machines) are mobile, represented by particles,
and the substrate and product are explicitly introduced as particles.

% If a substrate particle is in contact with a machine,
We assumed that only those particles in contact with each other can react,
and, as long as they are in contact, the reaction rate does not depend on the distance.
Each enzyme in a resting state starts the operation cycle
at rate $\alpha$ per substrate particle in contact with the enzyme.
Moreover, each enzyme in the resting state without a regulatory product
binds the product at rate $\beta$ per product particle in contact with the enzyme.
The substrate or product is removed from the system when these events occur.
The regulatory product alters the cycle initiation rate $\alpha$;
$\alpha = \alpha_{1}$ or $\alpha_{0}$ is used for enzymes with or without the regulatory
product, respectively.
Note that the definition of $\alpha$ and $\beta$ differs from
Model 1 and \cite{Togashi2015}.

The regulatory product dissociates at rate $\kappa$,
and also immediately upon the initiation of the cycle.
Likewise, at $\phi_{i} = \phi_{p}$ in the cycle, a new product is released
from the enzyme.
Given these events, a product particle is placed
at distance $r_{i}$ (the machine radius) from the center of the machine
to a random direction.

The substrate stochastically flows into the system at rate $\iota_{S}$ per surface area.
The position of the new substrate particle is determined randomly
with the minimum center-to-center distance $d_{min} = 0.1$. (It mimics
the situation whereby the substrate comes from out of the simulated 2D plane,
e.g., from cytosol to the plasma membrane.)
The substrate and product are also removed stochastically at rate
$\gamma_{S}$ and $\gamma$, respectively.

To consider crowding effects, in some cases, we additionally included
inactive particles (crowders; I) that interact with other particles through
the repulsive forces (which may affect the machines' phase $\phi_{i}$ through the potential).
However, these do not bind or react with other particles.
The number of the inactive particles is fixed at $N_{I}$, assuming no inflow or outflow.

Parameters $L_{x} = L_{y} = 128$, $\beta = 1000$, $\kappa = 10$, $\iota_{S} = 1$, $\gamma_{S} = 1$, and $\gamma = 10$ were used.
We initially distributed the particles randomly with the minimum center-to-center
distance $d_{initmin} = 0.3$ (and this also applies to the machines).
To avoid numerical instability after the placements of particles
(both initially and during the simulations),
the change to the particle position $\mathbf{x_{i}}$ and phase $\phi_{i}$ in a step
were limited to $0.2$ (scaled in keeping the direction), and $0.05$, respectively.
$\Delta t$ should be set such that the frequency with this limiter acts is sufficiently low 
to avoid changing the overall behavior of the system.
Here, $\Delta t = 5 \times 10^{-5}$ was used.

\section{Results and discussion}

\subsection{Basic properties of Model 1}

\begin{figure}
\includegraphics[height=57.6mm]{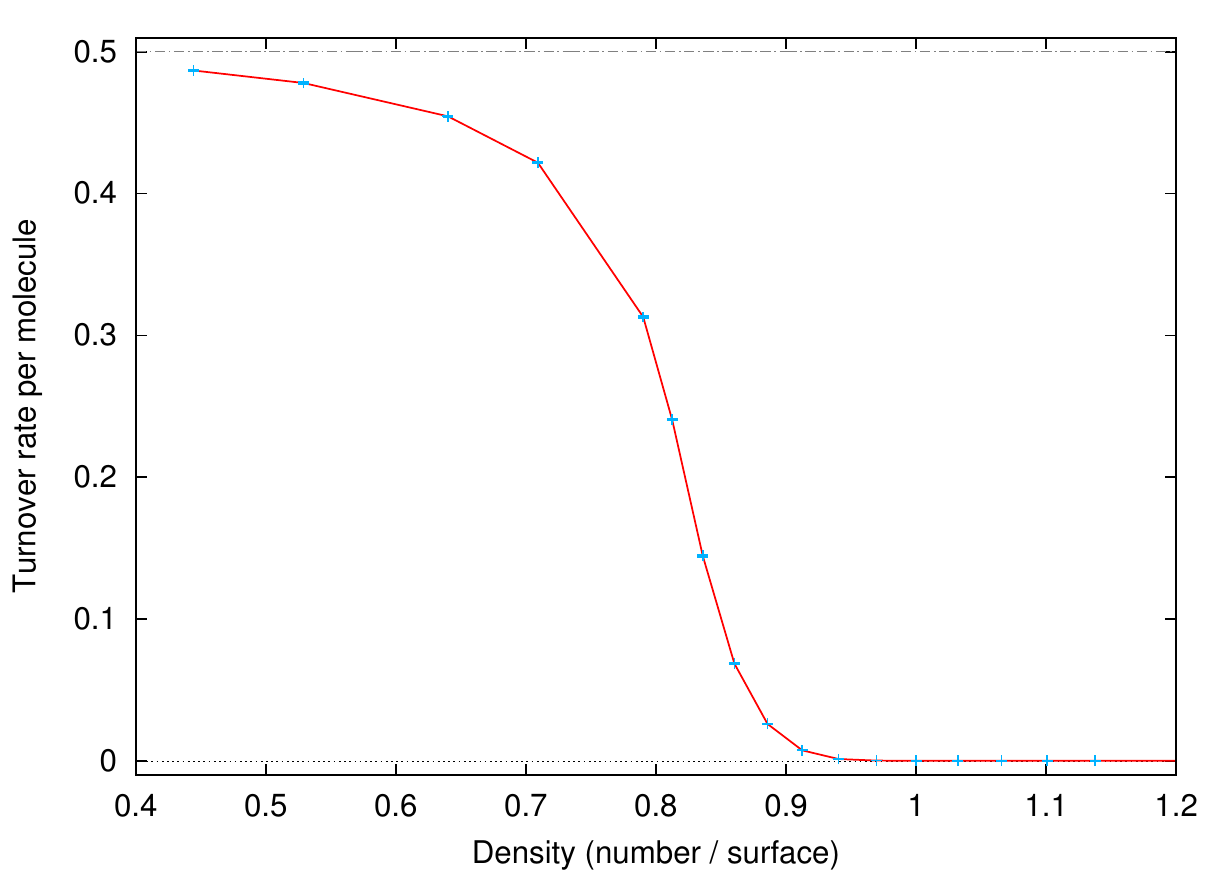}
\includegraphics[height=57.6mm]{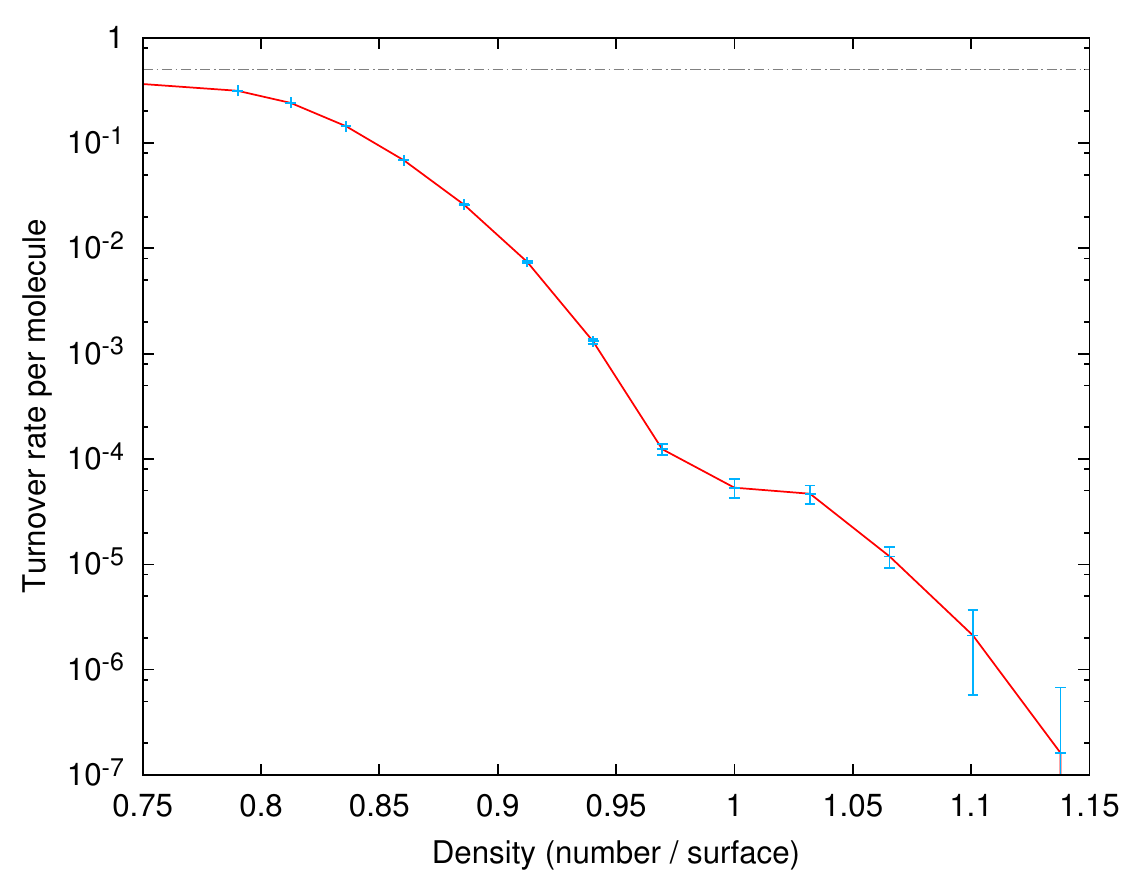}
\caption{Turnover rate against the density $N/(L_{x}L_{y})$ (number per surface area).
The reaction cycle cannot proceed because of the pressure
(repulsion from neighboring particles) when the density is very high.
The turnover rate sharply decreased around the number density 0.8,
showing phase transition to the stalled state.
Error bars show the mean $\pm$ standard deviation among the 10 trials.
}
\label{fig-turnover-density}
\end{figure}

\begin{figure}
\includegraphics[width=80mm]{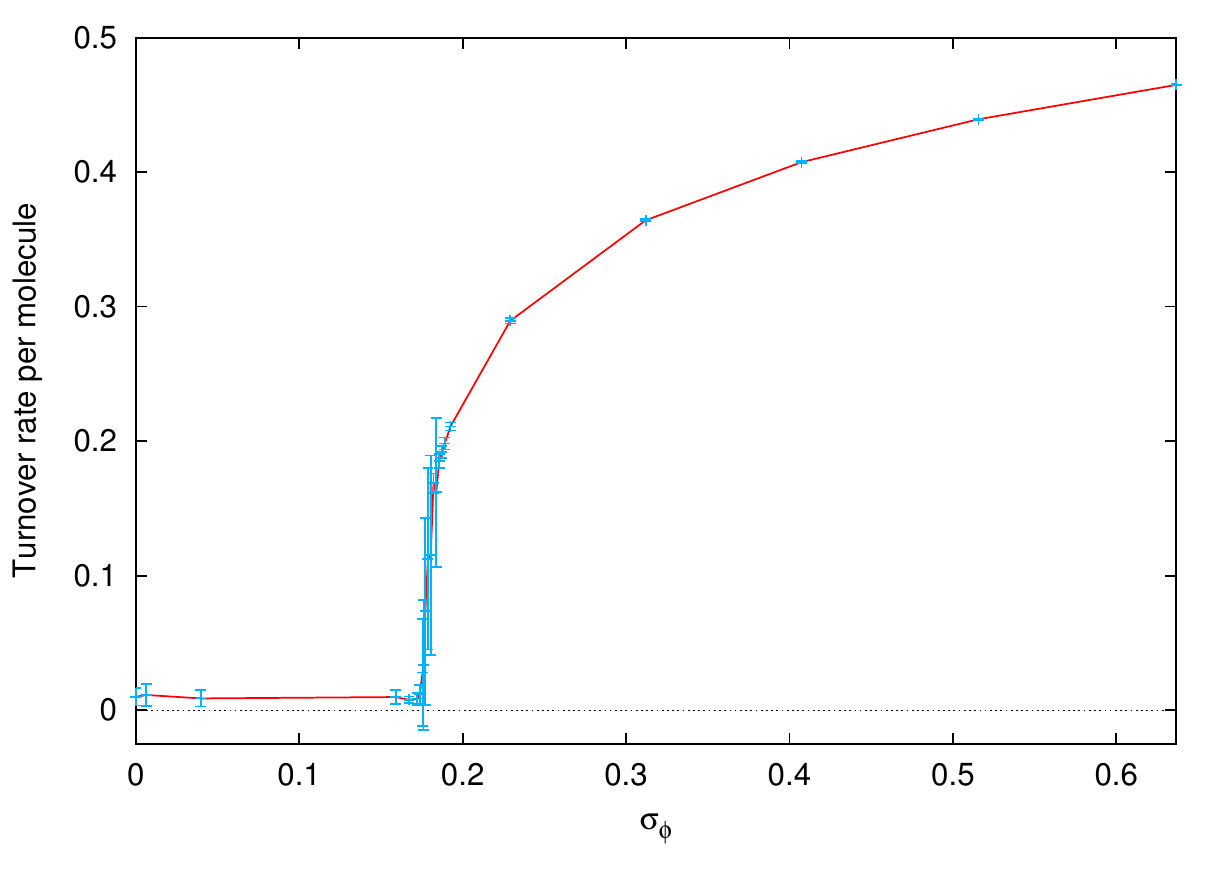}
\includegraphics[width=80mm]{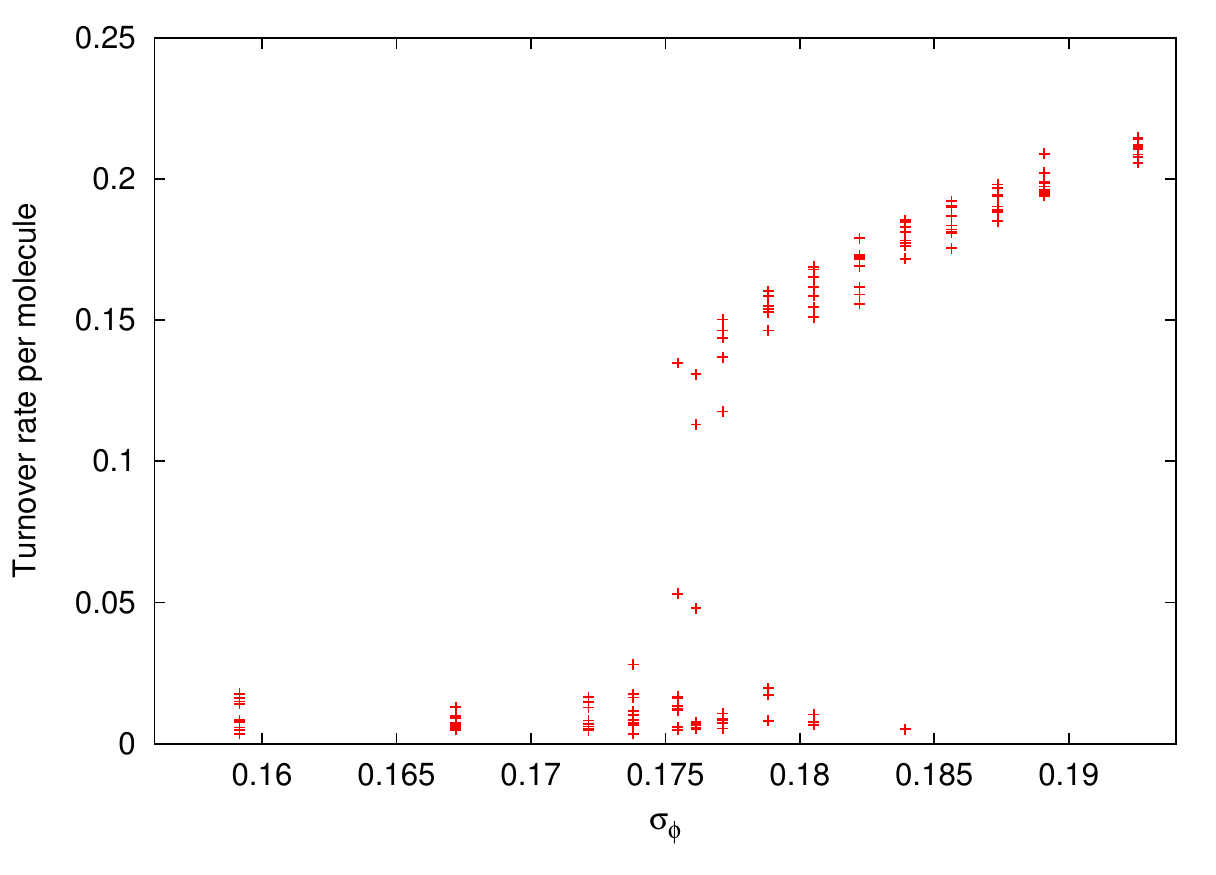}
\caption{Turnover rate against $\sigma_{\phi}$.
$\sigma=0$. 
When $\sigma_{\phi}$ is low, the system is in a stalled state.
As $\sigma_{\phi}$ increases, the fluctuations in the phase become so strong
that the phase $\phi_{i}$ can slip, and hence the reaction cycle can proceed
even at such a high density.
Steep increase in the turnover rate was observed around the transition point.
Error bars (top) show the mean $\pm$ standard deviation among the 10 trials.
Magnified view (bottom). The average turnover rate observed in each trial is shown.
}
\label{fig-turnover-sigmaphi}
\end{figure}

\begin{figure}
\includegraphics[width=80mm]{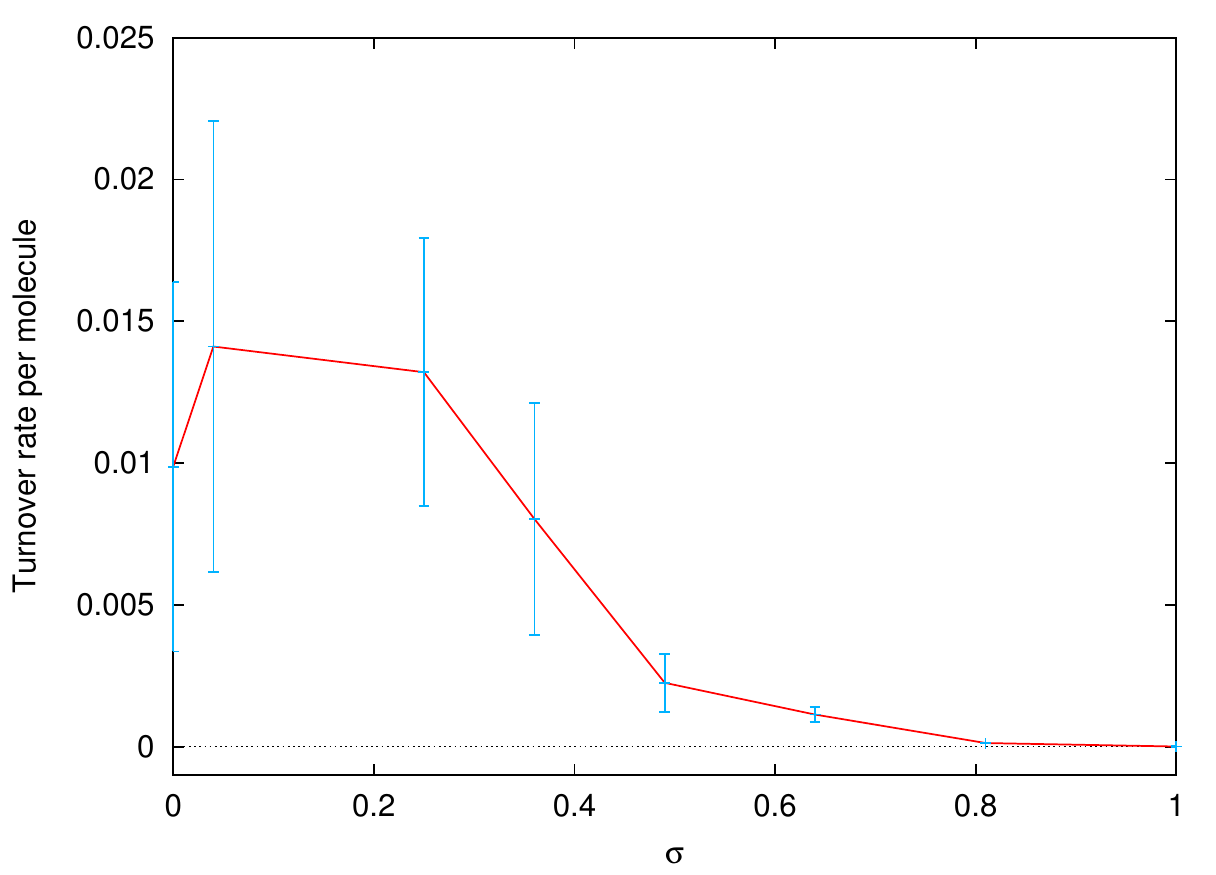}
\caption{Turnover rate against $\sigma$.
$\sigma_{\phi}=0$. The system is in a stalled state. Note that
the rate is quite low, reflecting a situation when only limited machines
can operate in the state.
At higher $\sigma$, the effective pressure on the machines from the surroundings
also becomes large, hindering the progress of the cycle.
Hence, in contrast to high $\sigma_{\phi}$ cases, the operations were blocked
even further.
Error bars show the mean $\pm$ standard deviation among the 10 trials.
}
\label{fig-turnover-sigma}
\end{figure}

We first studied the basic properties of Model 1 introduced in
Sections \ref{sect-common} and \ref{sect-model1}, particularly with regard to the overall
turnover rate of the machines.
As expected, the machines interfered with one another.
Specifically, when the radius increased (i.e., expanded),
the progress of the phase was suppressed when pressed by repulsive forces
from the particles in contact.
Consequently, the overall behavior depended on the density of particles.
We observed phase transition in the system between
an active state, where the machines can run their operation cycles, and
a stalled state, where the machines are tightly packed and the operation cycles
only rarely proceed.
As shown in Fig. \ref{fig-turnover-density}, the turnover rate steeply dropped
as the density increased, almost exponentially decreasing beyond that point.

Between these two states, another transition is caused by the fluctuations
in the phase, denoted by $\sigma_{\phi}$.
Figure \ref{fig-turnover-sigmaphi} shows the dependence of the turnover rate
on $\sigma_{\phi}$.
When the fluctuations became strong, the rate suddenly increased
around $\sigma_{\phi} = 0.176$.
Because of the strong fluctuations, the energetic barrier could be overcome,
such that the phases $\phi_{i}$ could slip and were distributed,
which enabled the operation cycle to proceed.
By contrast, fluctuations in positions (i.e. diffusion) represented by
$\sigma$ were affected differently.
As shown in Fig. \ref{fig-turnover-sigma},
at high $\sigma$, the turnover rate (already very low in the stalled state)
further decreased with an increase in $\sigma$,
which enhanced effective pressure from neighbors.

The transition here may appear similar to Tjhung and Berthier \cite{Tjhung2017},
who reported phase transition from slow to fast diffusion states
by increasing the ``activity'' (i.e., relative change of the particle radius).
However, in their model, the oscillation is always at the constant frequency
(corresponding to zero phase-susceptibility and an infinite cycle-initiation rate
case in our Model 1)
and the total area fraction of the particles is kept constant.
By contrast, the transition observed in our model is primarily
on the reaction turnover rate, caused by interference to the machine phase. That is, the oscillation itself stalls.
Combined with their discussion, our model suggests the possibility of
another kind of phase transition in the crowd of cells.

\begin{figure}
\includegraphics[width=53mm]{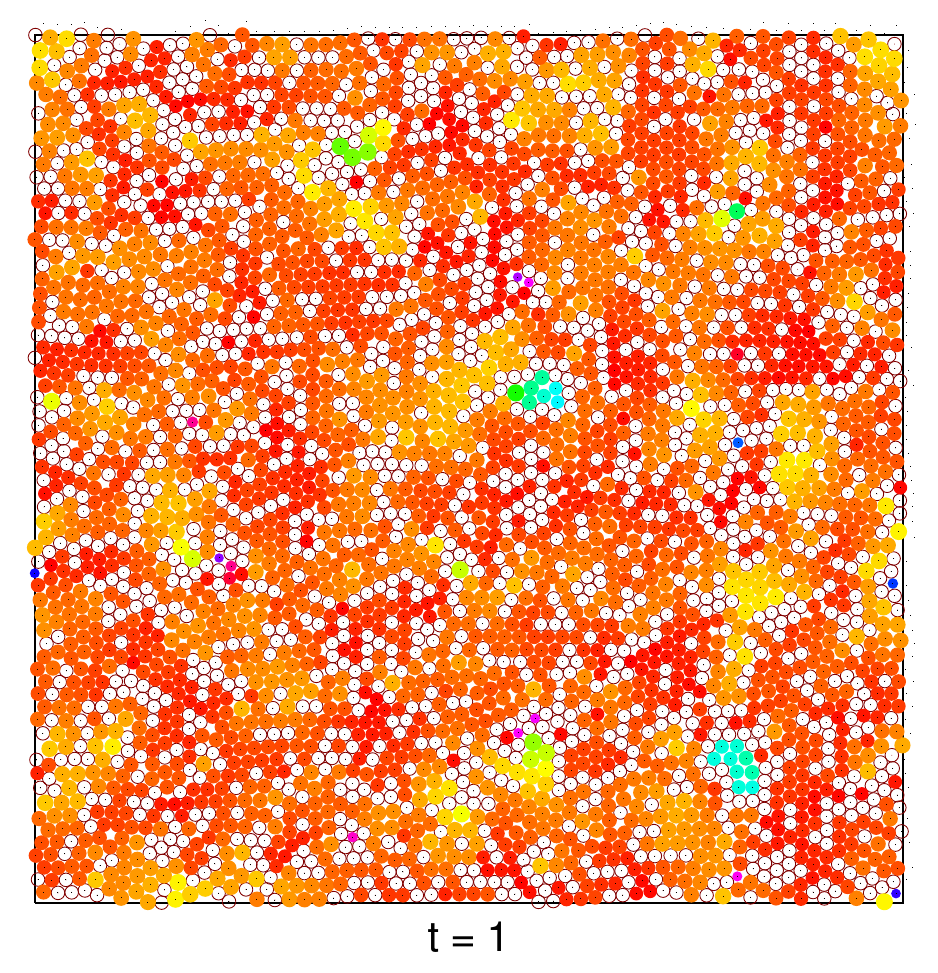}
\includegraphics[width=53mm]{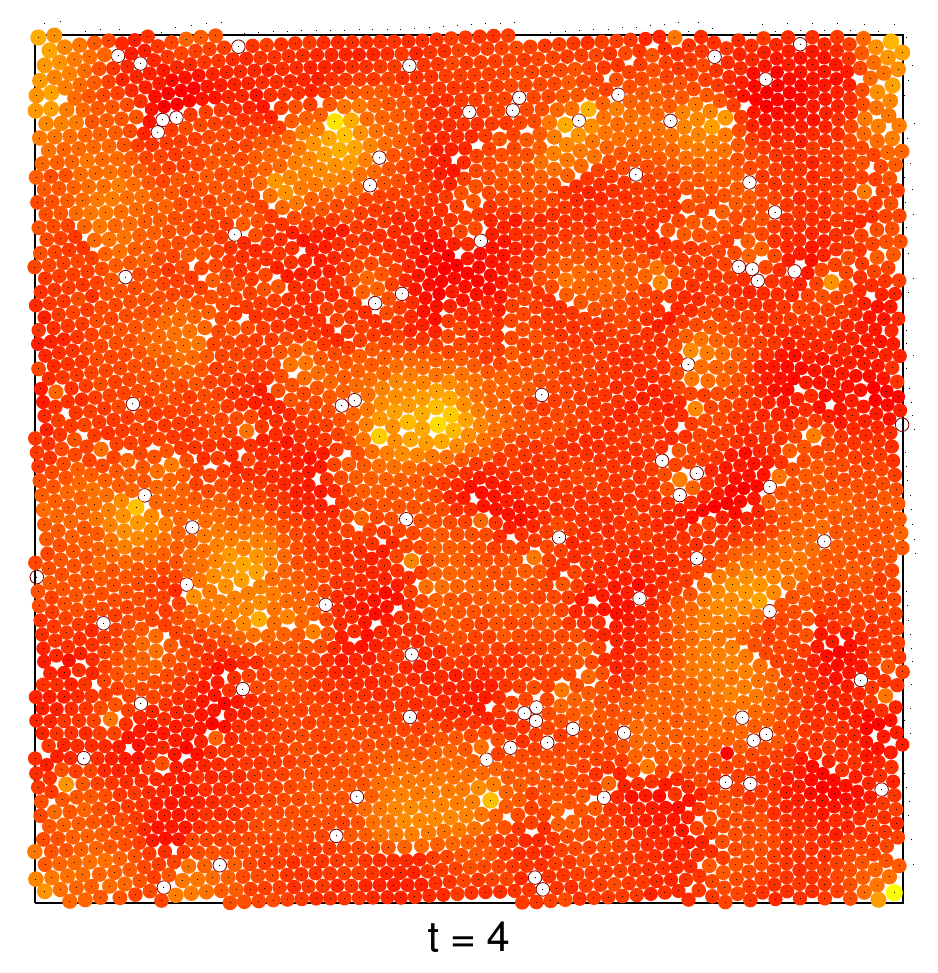}
\includegraphics[width=53mm]{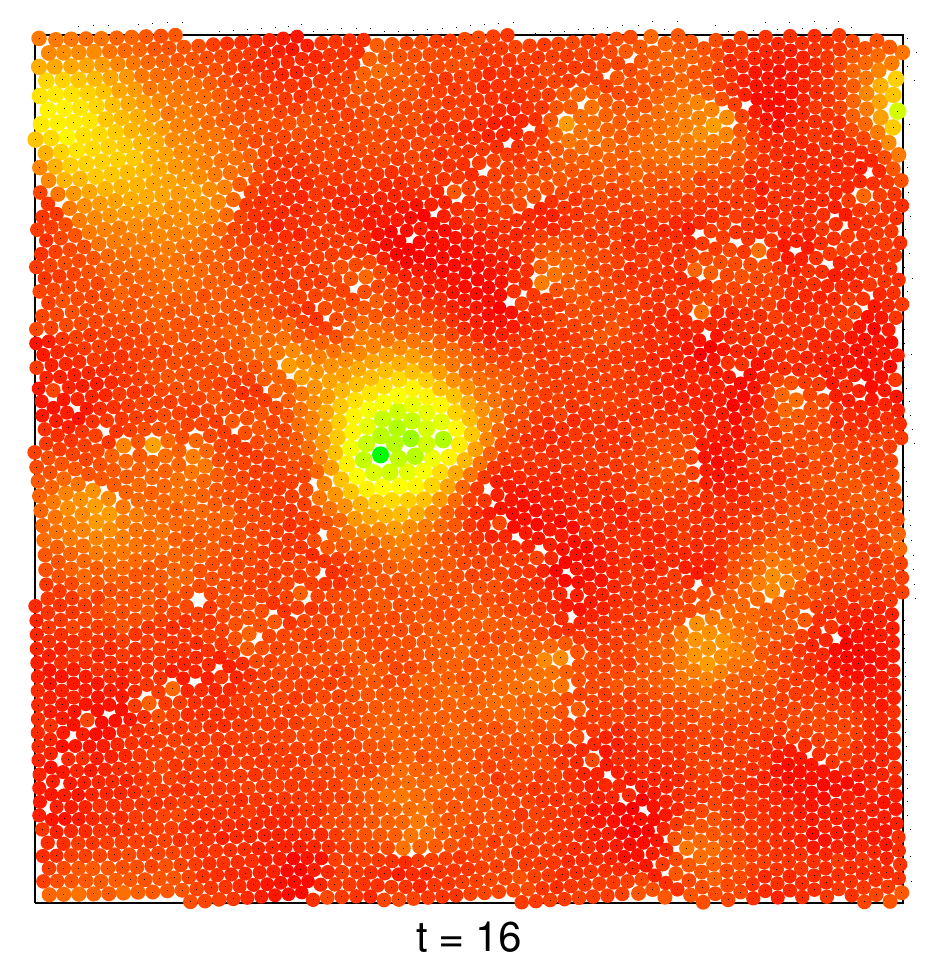}
\includegraphics[width=53mm]{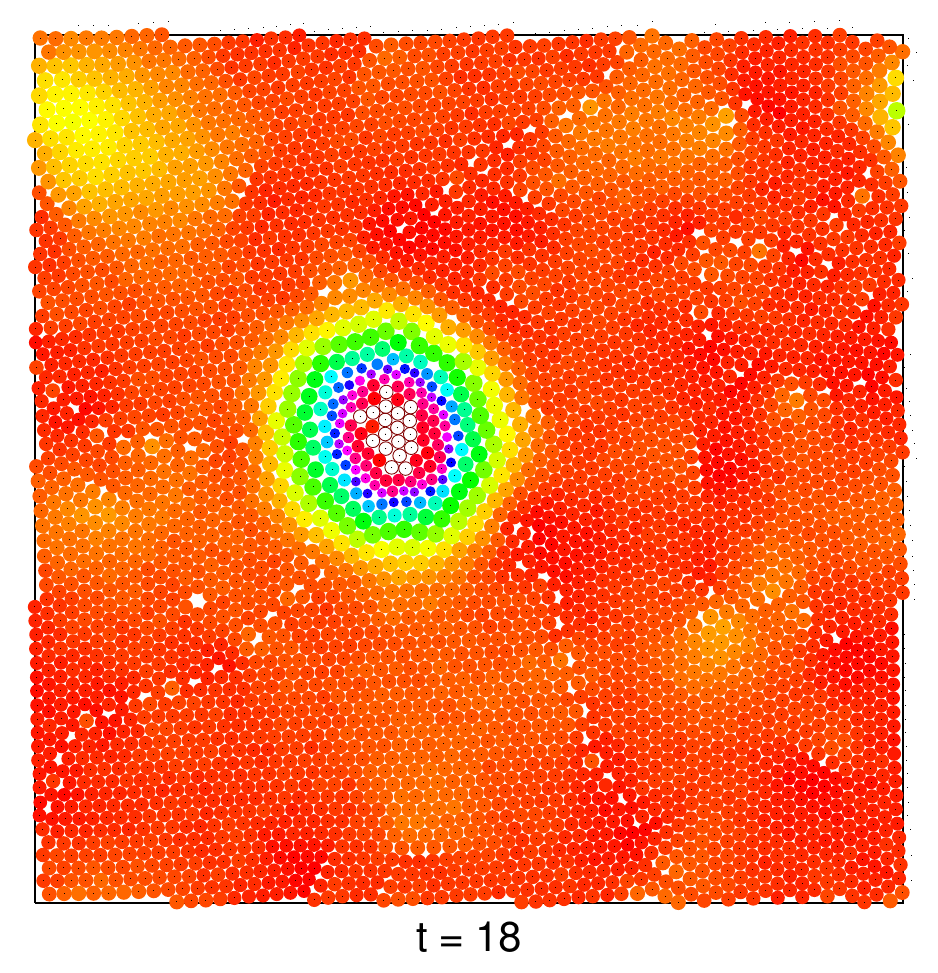}
\includegraphics[width=53mm]{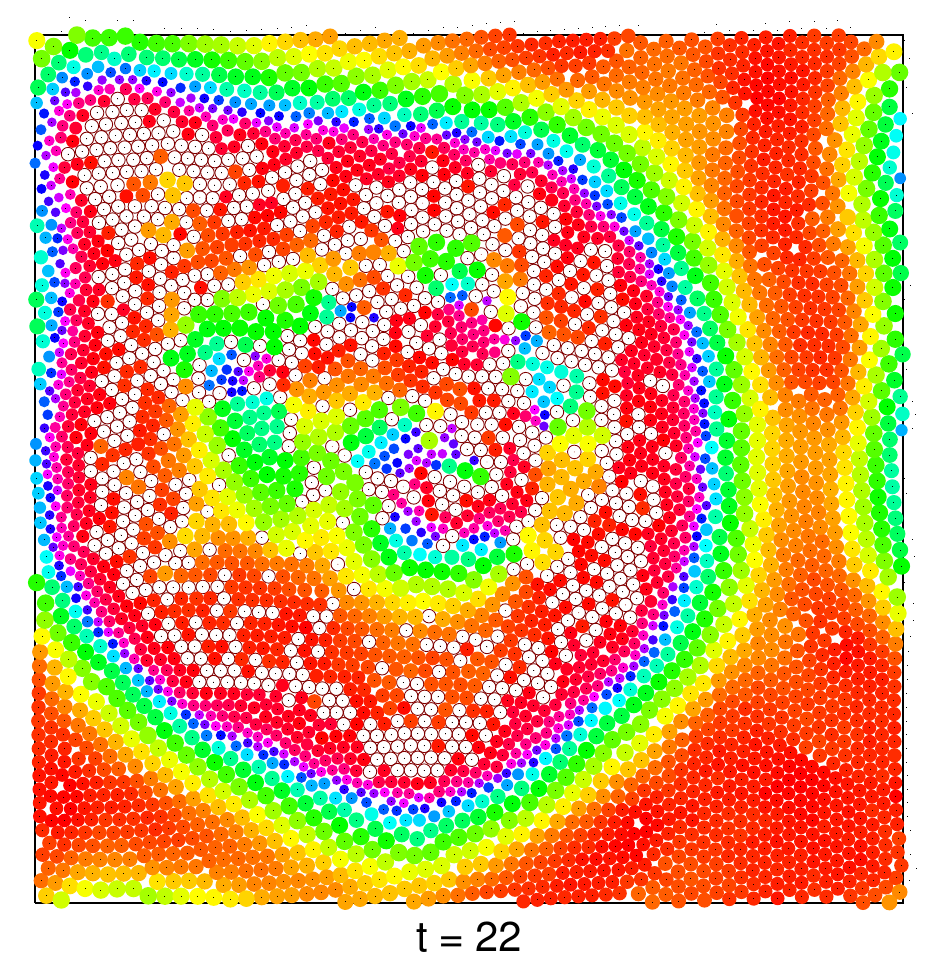}
\includegraphics[width=53mm]{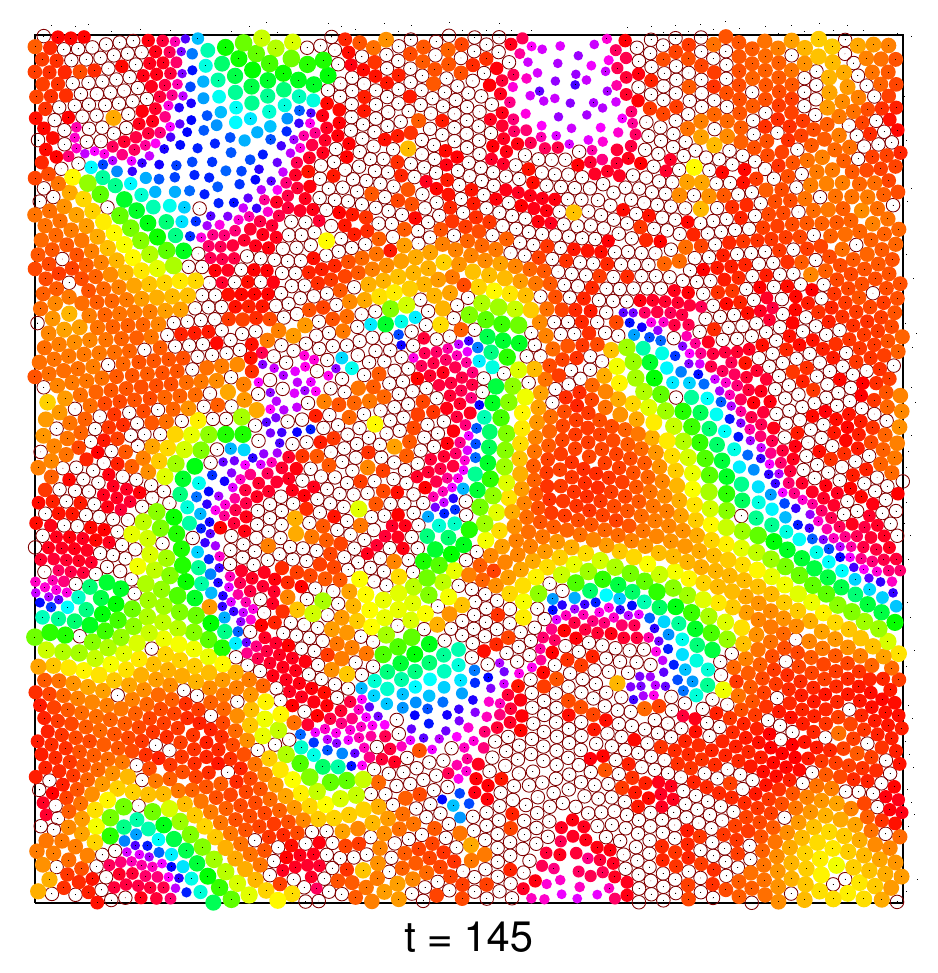}
\caption{Model 1 with additional phase coupling.
Snapshots at $t=1$, $4$, $16$, $18$, $22$, and $145$ are shown.
Kuramoto-type couplings between phase variables of machines in contact with each other
are introduced (see Section \ref{sect-phasecoupling} for details).
The neighboring machines tend to synchronize through the coupling.
As all the machines try to expand simultaneously,
they stack at the beginning of the cycle ($t=4$).
Then, a machine overcomes the barrier and starts to shrink ($t=16$),
which allows nearby machines to proceed.
}
\label{fig-snapshot-kuramoto}
\end{figure}

We next introduced additional couplings between $\phi_{i}$,
as described in Section \ref{sect-phasecoupling}.
With attractive interactions ($\epsilon > 0$ in Eq. (\ref{eqn-kuramoto})),
the phase variables $\phi_{i}$ tend to synchronize,
which may hinder the operation in a similar manner to the high density case above.
Figure \ref{fig-snapshot-kuramoto} shows snapshots from the simulation.
First, the machines were quickly synchronized. As a result,
they were stacked during the expansion of the radii at the early stage of the cycle ($t=4$).
Due to the fluctuations, by chance, a particle overcame the potential barrier ($t=16$);
the particle started to shrink, making space for nearby particles.
Then, the operation of the neighbors could proceed ($t=18$).
As a result, traveling waves were observed ($t=22$).

When particles shrunk, free space appeared around them, 
enhancing their diffusion.
Moreover, by losing contact with neighbors, phase coupling was also weakened,
allowing phase $\phi_{i}$ to disperse.
These effects resulted in fragmentation and disturbance of the wave patterns ($t=145$).

\subsection{Reaction-diffusion behavior of Model 2}

\begin{figure}
\includegraphics[width=80mm]{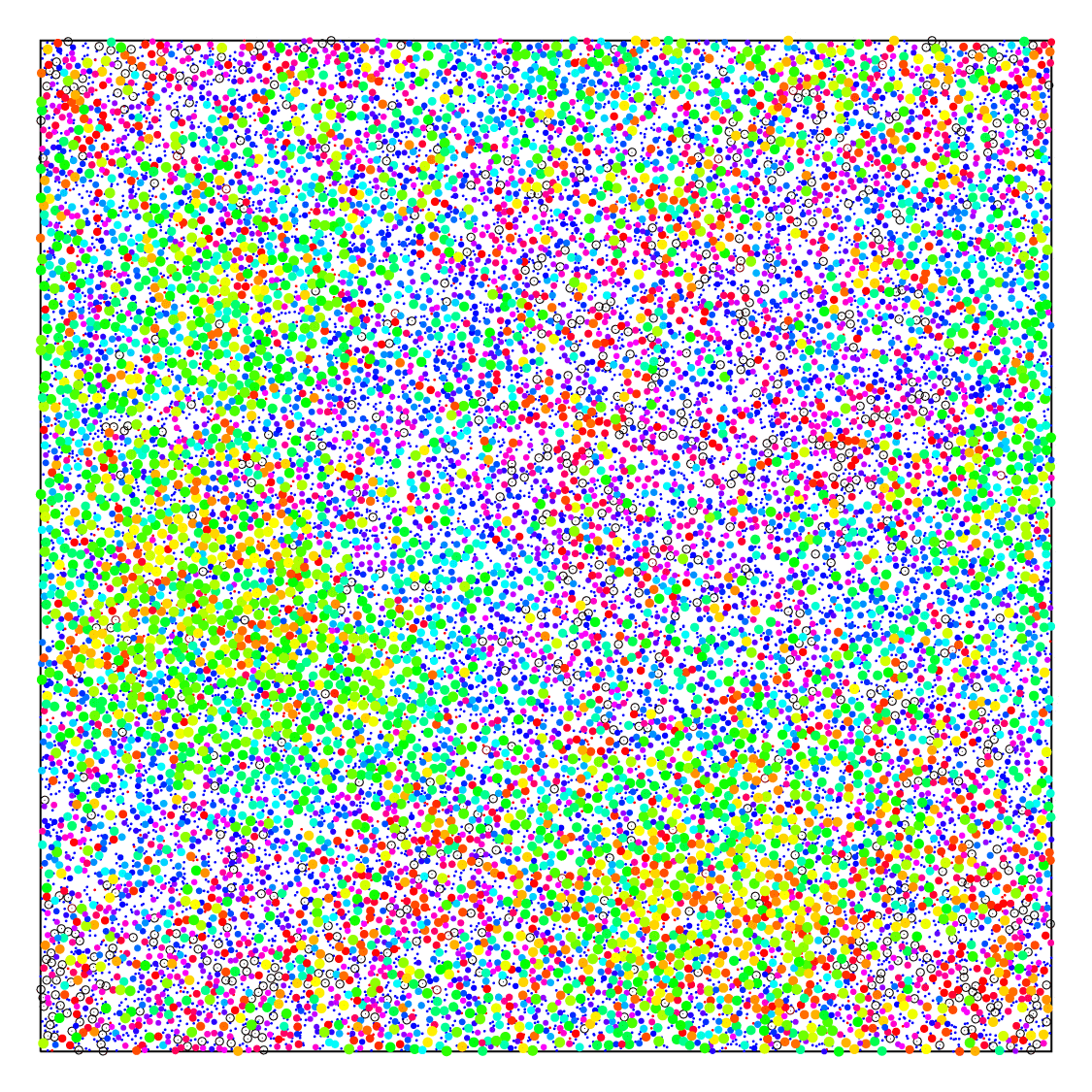}  % new21-mp0.001-vn0.1-pn0
\caption{Snapshot of Model 2 with fast diffusion, where $t=100$.
Parameters: $A=0.3$, $\phi_{p} = 0.2$, $\alpha_{0} = 1$, $\alpha_{1} = 1000$,
$\mu_{\phi} = 0.001$,
$\sigma = 10$, $\sigma_{\phi}=0$, $N = 10240$, and $N_{I} = 0$.
Traveling waves and spirals similar to those reported in \cite{Casagrande2007,Togashi2015} were observed.
}
\label{fig-snapshot-waves}
\end{figure}

\begin{figure}
\includegraphics[width=80mm]{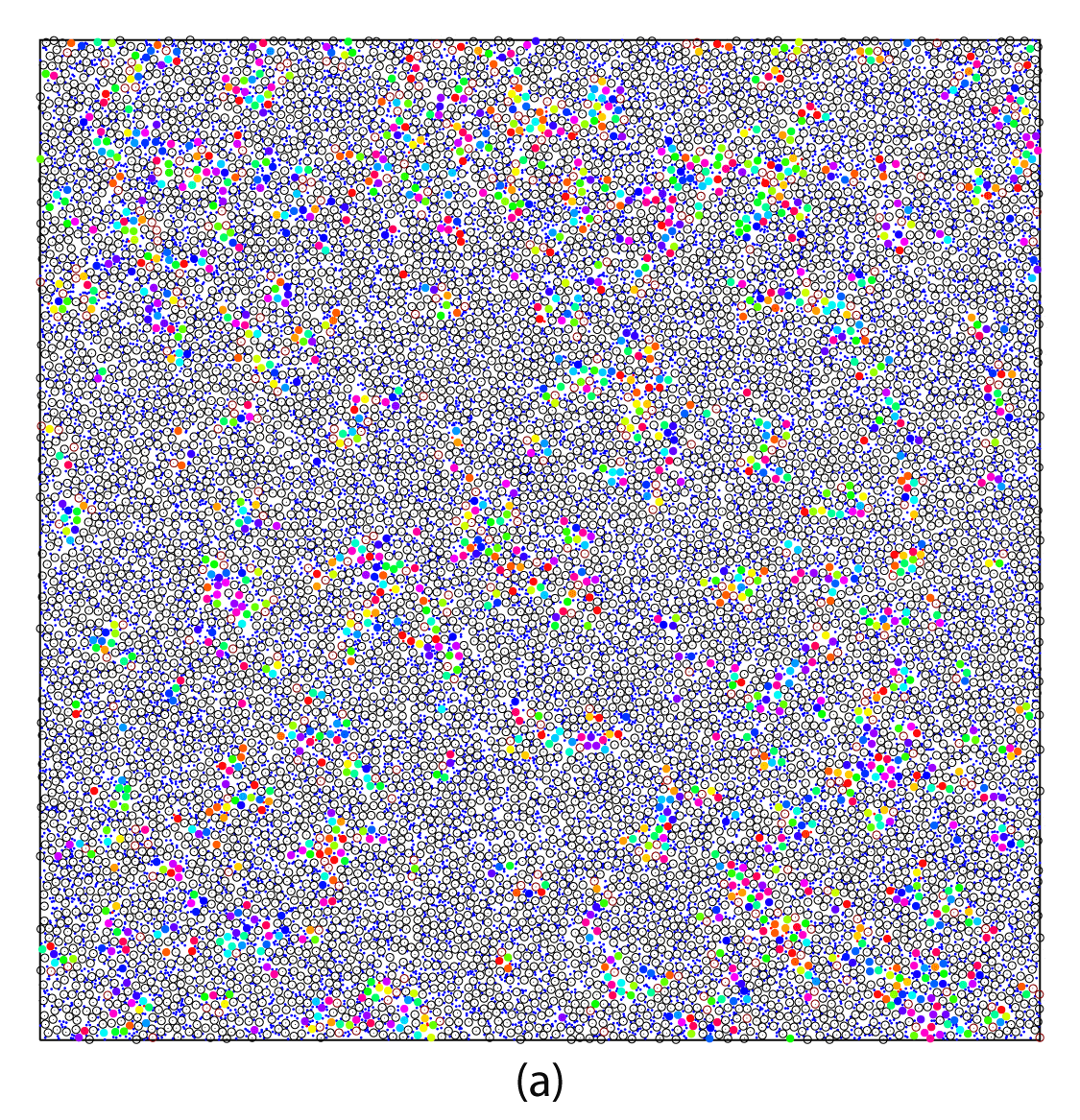}
\includegraphics[width=80mm]{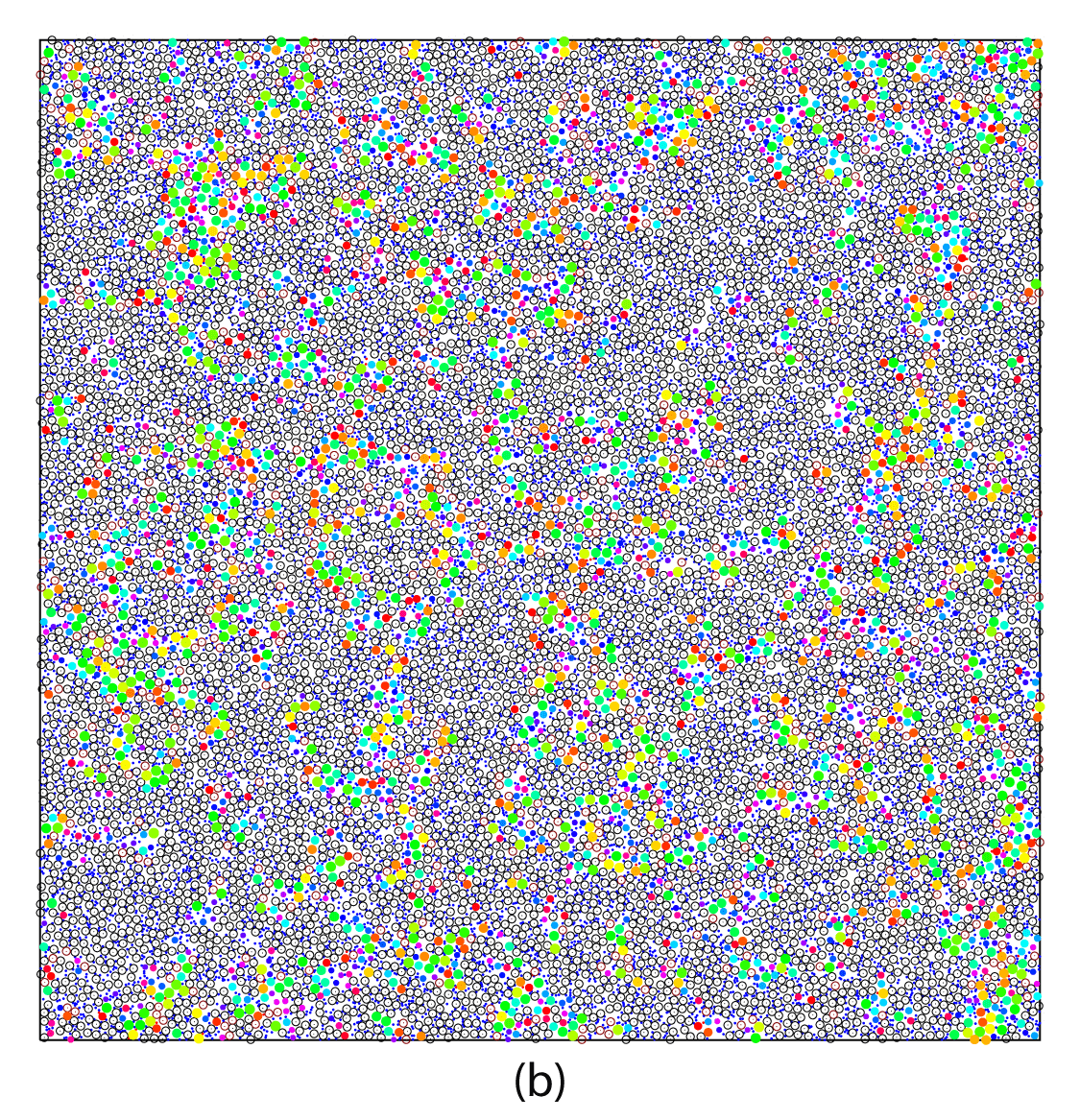}
\includegraphics[width=80mm]{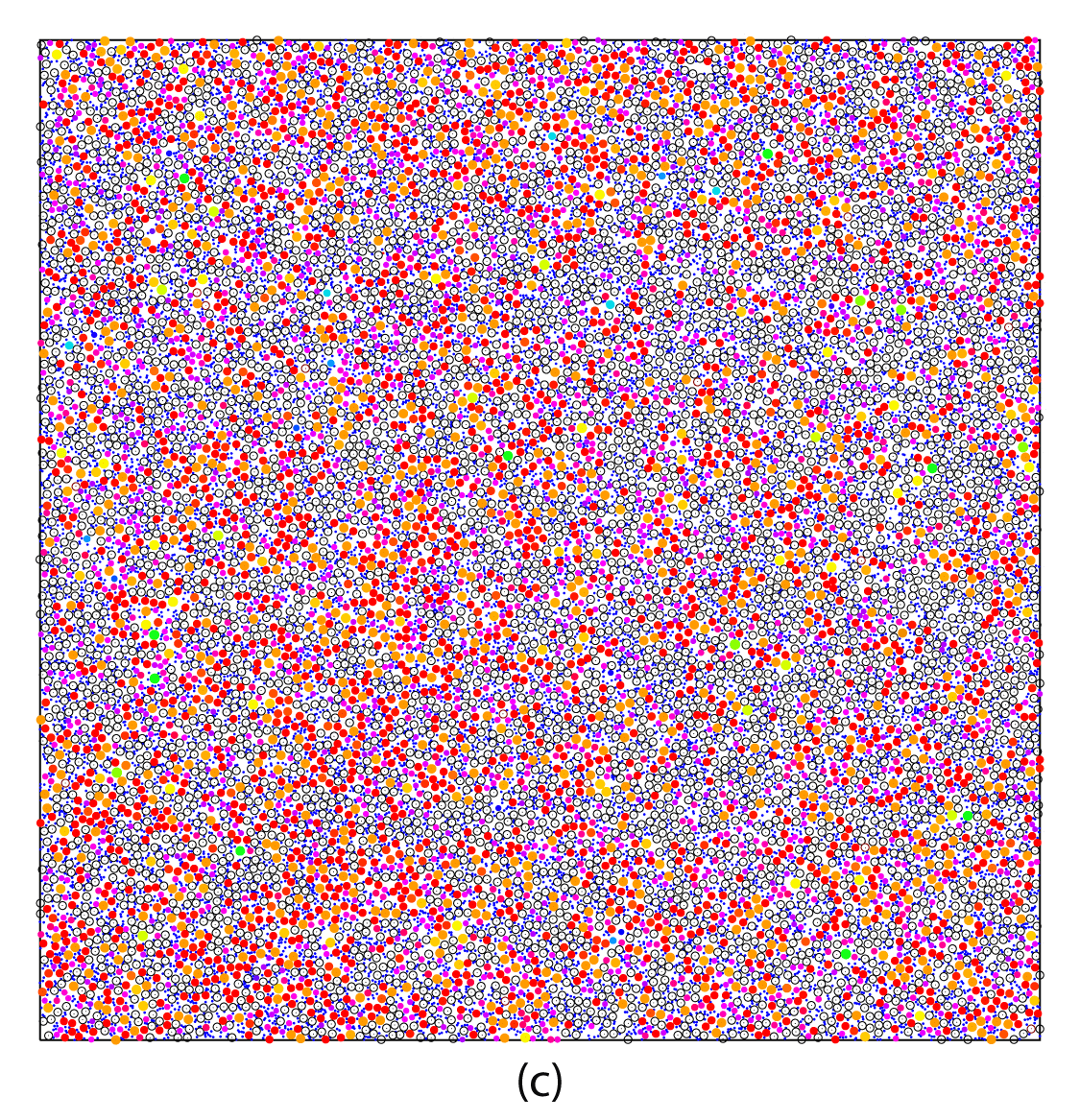}
\includegraphics[width=80mm]{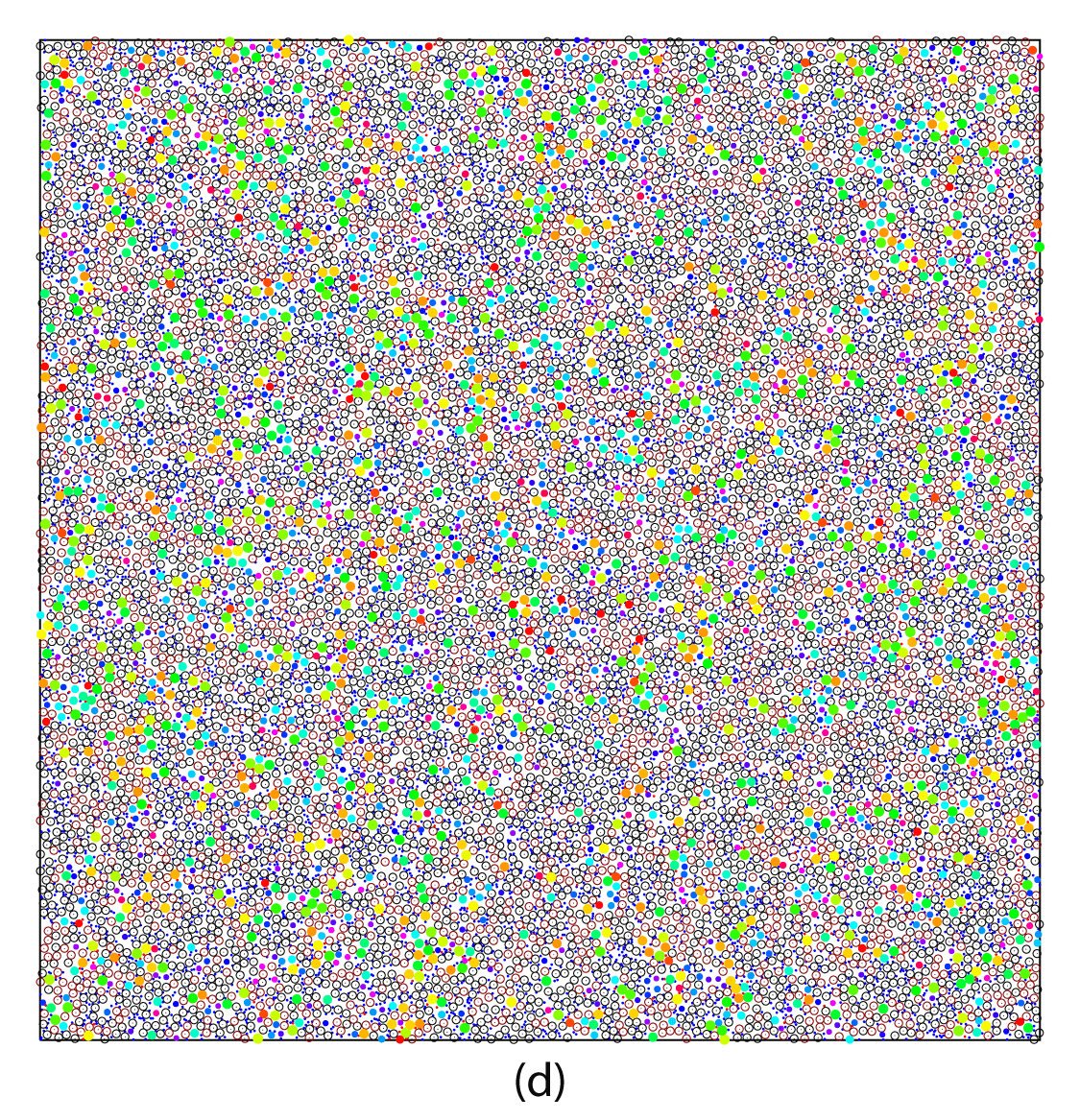}
\caption{Snapshots of Model 2 with slow diffusion, where $t=100$.
Parameters: $\phi_{p} = 0.1$, $\alpha_{0} = 0.01$, $\alpha_{1} = 10$,
$\sigma = 0.1$, $N = 10240$, and $N_{I} = 0$.
(a) $A=0$; (b) $A=0.3$, $\mu_{\phi}=0$, $\sigma_{\phi}=0$;
(c) $A=0.3$, $\mu_{\phi}=1$, $\sigma_{\phi}=0$; and (d) $A=0.3$, $\mu_{\phi}=1$, $\sigma_{\phi}=1$.
The system was segregated into active paths and inactive clusters (a,b).
With high $\mu_{\phi}$, the operation was blocked by the pressure from neighbors (c),
which could be resumed by large $\sigma_{\phi}$ (d), in the same way as in Model 1.
}
\label{fig-snapshot-paths}
\end{figure}

\begin{figure}
\includegraphics[width=80mm]{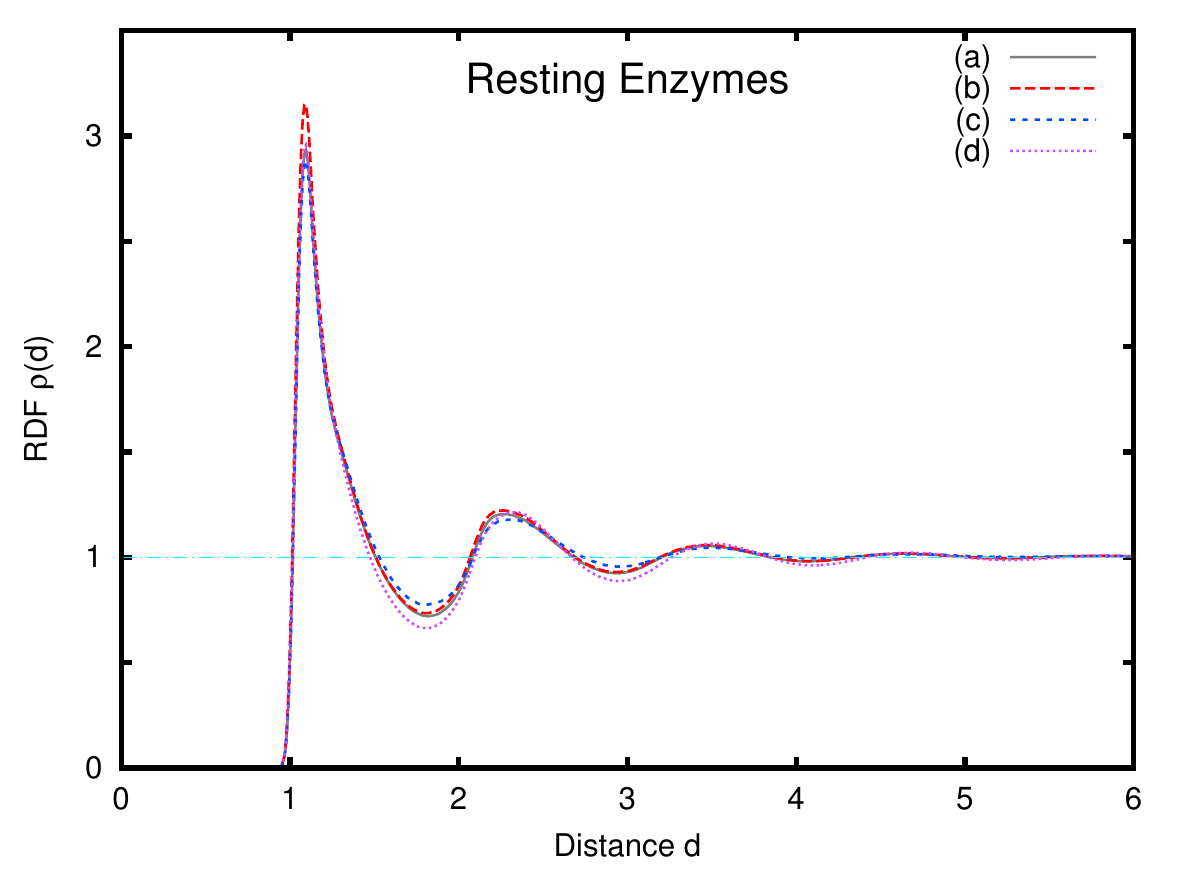}
\includegraphics[width=80mm]{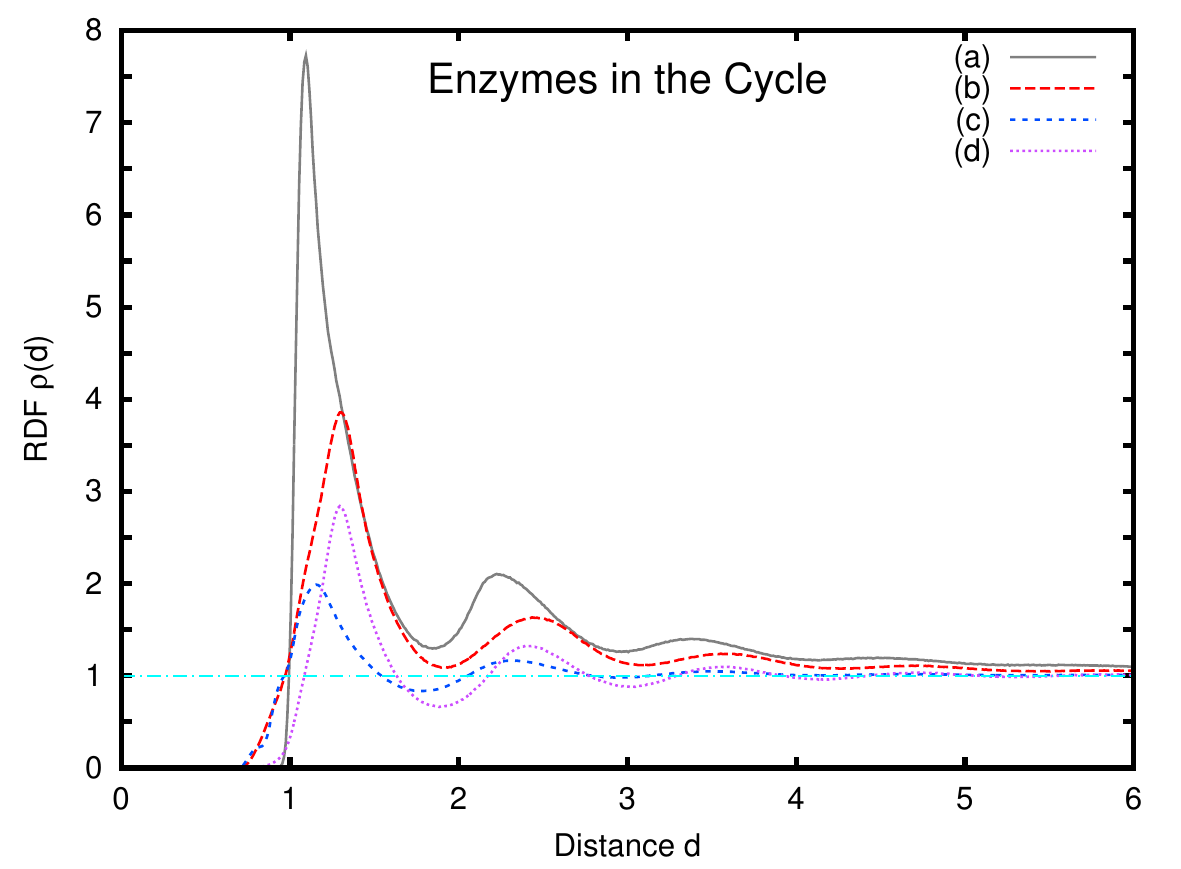}\\
\includegraphics[width=80mm]{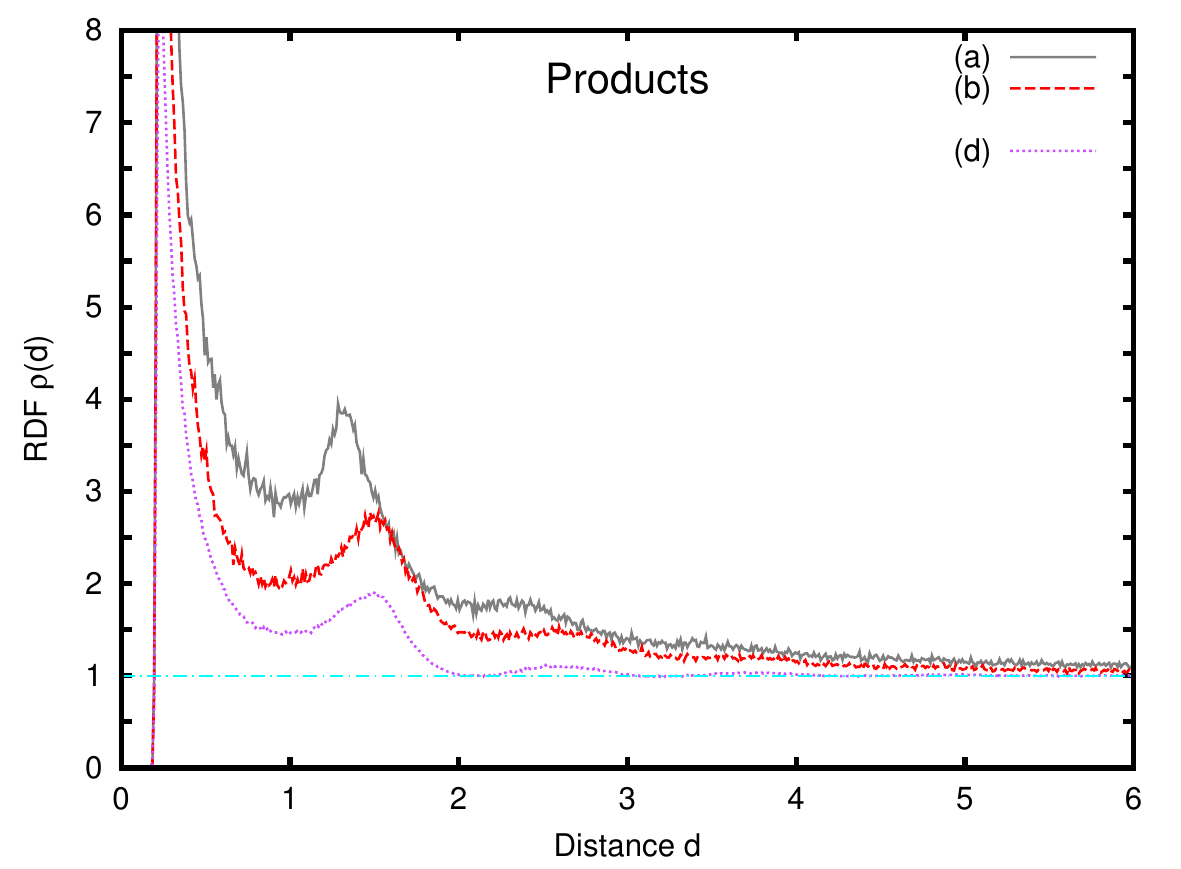}
\includegraphics[width=80mm]{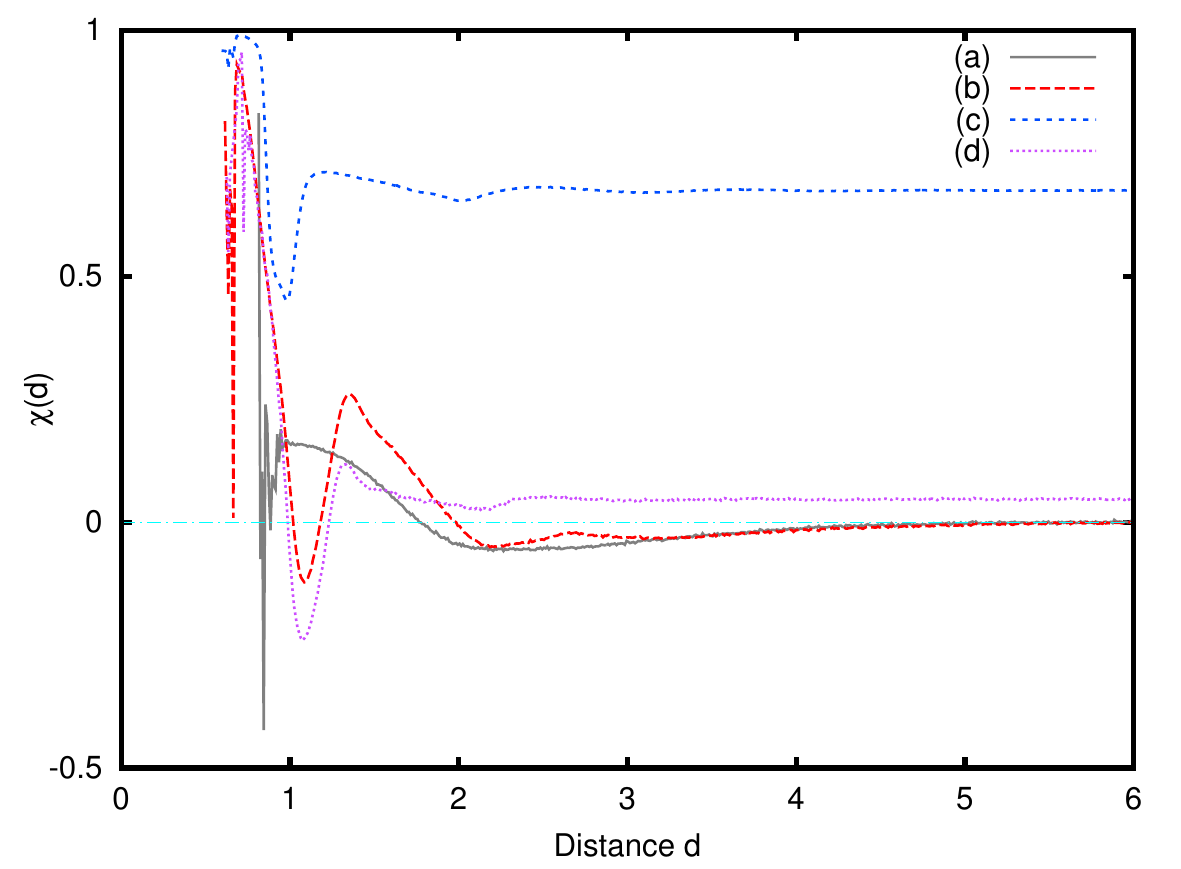}
\caption{Radial distribution functions $\rho (d)$ of
enzymes in a resting state (top left) and in the cycle (top right), and products (bottom left).
Phase correlation $\chi (d)$ (bottom right) is also shown.
The parameter sets (a)--(d) are the same as in Fig. \ref{fig-snapshot-paths}.
Note that there was no product in case (c).
When $A=0$ (a), $\rho (d)$ of enzymes shows a sharp peak around their diameter $d \sim 2r_{0} = 1$ and
additional peaks near integral multiples of that, reflecting nearly close packing of these particles.
$\rho (d)$ of resting enzymes is almost the same for the other cases (b)--(d).
In contrast, 
$\rho (d)$ of enzymes in the cycle in case (a) is significantly positive within a range $d \lesssim 3$,
showing localization of enzymes in the cycle, i.e., localized operation.
$\rho (d)$ of products also has such sharp peaks.
Its decay with distance also shows localization of products within $d \lesssim 2$.
The peak at distance slightly larger than $1$ represents localization of products around enzymes.
Phase correlation $\chi (d)$ is positive (i.e. in phase) for the nearest neighbor $d \sim 1$.
It is slightly negative for the second nearest ($d \sim 2$) and further,
where the communication is delayed.
In case (b) with $A=0.3$, the peaks in $\rho (d)$ of enzymes in the cycle are broadened
because of the variable radii; products are dispersed by the enzyme motion as well.
$\chi (d)$ shows a steep trough around $1$, which implies that centers of enzyme particles
cannot come so close unless they are operating antiphase.
With large $\mu_{\phi} = 1$ (c), the enzymes were completely stacked up in the cycle,
resulting in highly positive $\chi (d)$ regardless of distance.
When $\sigma_{\phi}$ was also increased to $1$ (d), the operation was possible again.
However, the localization of machine operation and products was suppressed.
}
\label{fig-RDF}
\end{figure}

\begin{figure}
\includegraphics[width=80mm]{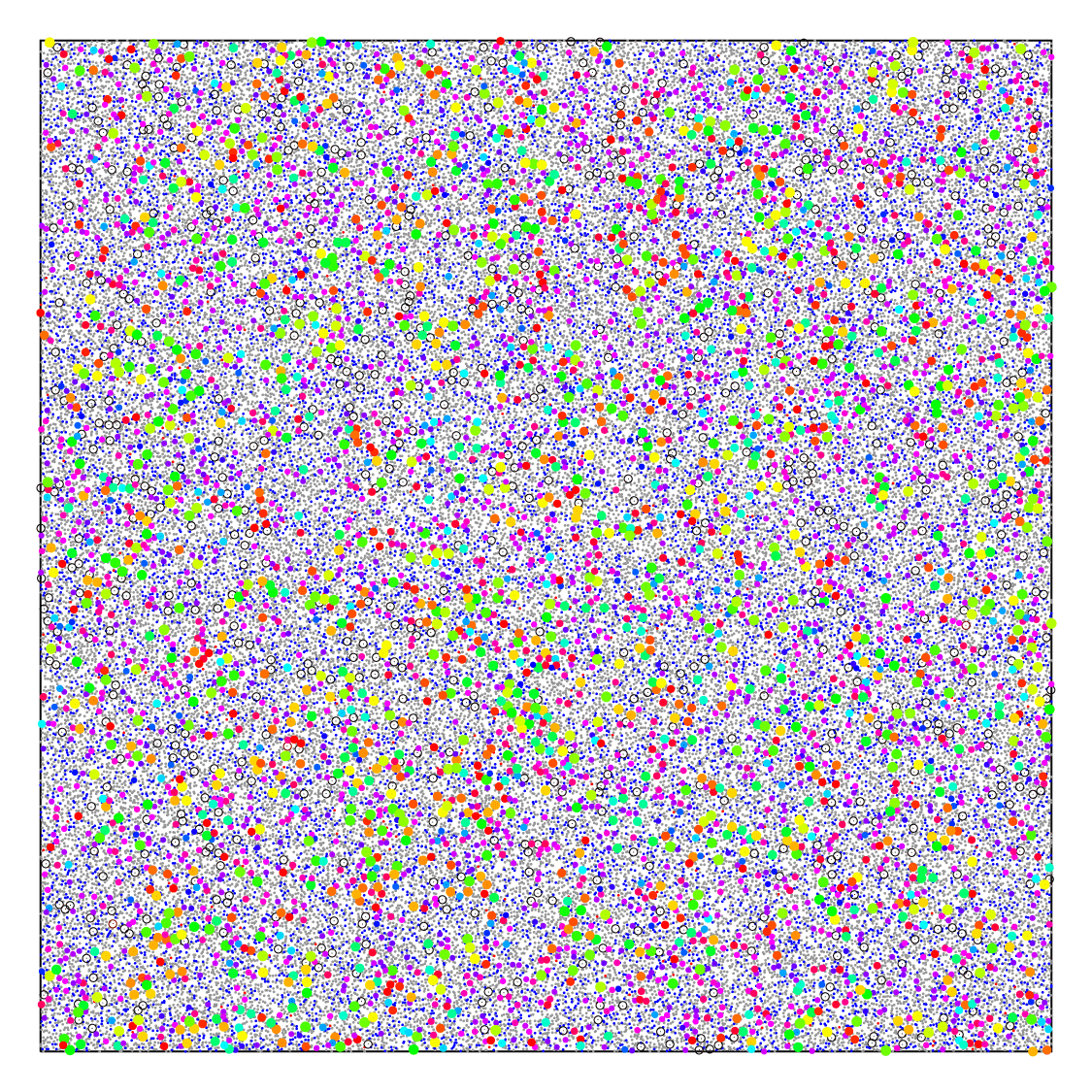}
\caption{Snapshot of Model 2 with inactive particles, where $t=100$.
Parameters: $A=0.3$, $\phi_{p} = 0.2$, $\alpha_{0} = 1$, $\alpha_{1} = 1000$,
$\mu_{\phi} = 0.0001$,
$\sigma = 10$, $\sigma_{\phi}=0$, $N = 5120$, and $N_{I} = 51200$.
}
\label{fig-snapshot-inactive}
\end{figure}

We next simulated Model 2, as described in
Sections \ref{sect-common} and \ref{sect-model2}.
This model is a combination of the particle model and
the enzymatic reaction model introduced in \cite{Togashi2015}.
We focused on spatiotemporal patterns
and compared our results with those of the previous study, to consider the effects
of excluded volume and its interference with phase $\phi_{i}$.

When the density is low, or the diffusion is sufficiently fast,
the behavior is expected to be similar to that shown by our previous model.
Indeed, we observed traveling waves and spiral-like patterns
in the current model, as shown in Fig. \ref{fig-snapshot-waves}.

Since the enzymes interact with each other through the product molecules
(i.e., positive feedback by allosteric activation),
the behavior is expected to change when the diffusion of the product
is limited by crowding.
Figure \ref{fig-snapshot-paths} shows a system under such a condition.
There are clusters of enzymes in a resting state without the regulatory
product (gray circles).
Once a cluster of such resting enzymes forms,
the diffusion inside the cluster is hindered by crowding,
making it difficult for product particles to penetrate into the cluster
(i.e., crowding out).
Because the spontaneous reaction rate without the regulatory product
(denoted by $\alpha_{0}$) is set low,
they tend to remain in that state,
which results in relatively stable and inactive clusters.
By contrast, once the enzymes start the operation,
the change of the radii (for $A > 0$) makes space around the machine,
facilitating the diffusion of particles and
further enhancing the positive feedback by the product to nearby enzymes.
Such domains appear as flow-like ``paths'' of active enzymes and products.

As a result, segregation into active and inactive domains is observed.
In the former, reactions occur frequently, and this, in turn, facilitates
the diffusion of S and P and enhances further reactions
(i.e., positive feedback). In the latter, the machines are confined into blocks
and the penetration of P into those blocks is hindered.
To characterize the behavior, we plotted the radial distribution function
$\rho (d)$ of the enzymes and products and phase correlation $\chi (d)$ in
Fig. \ref{fig-RDF} (detailed descriptions are in the figure caption).
Radial distribution function $\rho (d)$ of enzymes in the cycle
indicates localization of machine operation within a distance range $d \lesssim 3$.
$\rho (d)$ of products also show their localization.
They are dispersed by the enzyme motion when $A > 0$.
Of course, the regions are not fixed, and change over time.
A transition similar to that in Model 1 was also observed for the change
of $\mu_{\phi}$ and $\sigma_{\phi}$ (Fig. \ref{fig-snapshot-paths} (c,d)).
When $\sigma_{\phi}$ was large, the operation was possible even for large $\mu_{\phi}$,
however the localization was lost.

The crowding effect can be also modulated by inactive particles,
which are not involved in the reaction (see Fig. \ref{fig-snapshot-inactive}
for an example).
Although only repulsive forces act between the particles
(cf. \cite{Awazu2007}),
spontaneous clustering of particles is sometimes observed,
because of the Asakura--Oosawa depletion forces \cite{Asakura1954,Binder2014}
mediated by the smaller particles (S, P, or I).
In this model, the radii of the machine particles change
depending on the states.
Hence, the effective interactions are modulated by the reactions.
In turn, the clustering may alter the rate of reactions there,
through, e.g., localization of P (positive regulation)
or blocking of P or S (negative regulation).

\subsection{Relevance to biological systems and applications}
\label{sect-relevance}

There is already considerable research on crowding effects
in the context of intracellular processes \cite{Elcock2010,Mourao2014,Rivas2016}.
In addition to these, our study offers insight into the interplay
between active conformational changes of molecular machines
and such crowded environments that include other machines.
The effect of a crowded environment on protein structures
and their stability has been experimentally investigated
(not always compaction, e.g., \cite{Banks2018}),
and those studies offer hints for further improvements to the model.

If the density is relatively low,
we can ignore the mechanical interactions between machines.
However, the actual environment in the cell is far from that.
Even glass-like properties of bacterial cytoplasm and its metabolism-dependent fluidization 
were reported \cite{Parry2014}. This strongly suggests the importance of
molecular machine activities (although this case is different from active diffusion
induced by cytoskeletal motors \cite{Brangwynne2009}).
The shape of crowders can be also important for diffusion \cite{Balbo2013}.

In reality, there are a huge variety of macromolecules in the cell.
We have considered only a single type of machines in each model thus far,
however it is easy to include multiple species of machines and different reaction cycles
(cf. \cite{Lerch2002}) into the model.
Since the formulation of the mechanical interactions is simple,
it is also easy to extend the model to 3-dimensional systems with spherical particles,
which is useful for modeling of complex and dynamic 3D structures such as chromatin
in the cell nucleus.
With such extensions, although in a very simplified way,
this model may serve as a framework to study the interplay between
the mechanical stress/strain network and the chemical reaction network in the cell.

In this model, the shapes of the machines are always isotropic (i.e., circular).
We can consider asymmetric changes to the shape, which may result in
the swimming-like behavior of the machines \cite{Huang2012},
and this is comparable to structurally resolved models of machines \cite{Togashi2007,Huang2013}.
The importance of the shape (i.e., deformation) for self-propelled particles was 
pointed out previously \cite{Ohta2009}.
Moreover, previous research reported that machines represented by force dipoles may
enhance diffusion and the transport of molecules through hydrodynamic effects
\cite{Mikhailov2006,Buyl2013,Mikhailov2015,Koyano2016} (which may also occur in the current case, as the surroundings
of the machine are not isotropic).
It would be interesting to realize active transport of the substrate or product
by the machine operation,
such that this could be compared to transport by active diffusion \cite{Brangwynne2009}.

In a model of molecular machines such as enzymes,
the internal state of the machine is also relevant to dynamic disorder
\cite{Zwanzig1990,Lerch2002,Lerch2005,Terentyeva2012},
although the state is represented exclusively by a single variable in the current model.
In a more general sense, the machine cycle denoted by phase $\phi_{i}$ represents
the delay in the operation.
Beyond crowds of nanomachines such as enzymes,
the model is applicable to crowds of cells,
with feedback delays at the level of cellular processes such as gene expression---which is sometimes modeled as delayed reaction-diffusion systems \cite{Seirin2012,Seirin2015}.
The mobility and fluctuations (denoted by $\mu_{i}$, $\mu_{\phi}$, $\sigma$ and $\sigma_{\phi}$, respectively)
may depend on the scale of the objects.
The model is also convenient for cells with intrinsically oscillatory
or excitable dynamics, such as cellular slime molds \cite{Arai2010,Gregor2010,Mehta2010}.
Moreover, the mechanical interference between machines could be interpreted in the context of
mechanochemical coupling in tissues \cite{Naganathan2017}.
It may be further adapted to more macroscopic oscillating droplets
and self-propelled objects \cite{Toiya2010,Thutupalli2013,Suematsu2018}
or artificial micro-machines.

If the system includes positive feedback (such as the regulatory product in Model 2),
effects from small numbers of molecules may appear, as observed in autocatalytic
reaction systems \cite{Togashi2001,Saito2015,Nakagawa2016,Shnerb2000,Togashi2004}.
In the current model, a single machine that runs the cycle ahead may facilitate
the operation of others through mechanical interference,
as seen in Fig. \ref{fig-snapshot-kuramoto}.
This may also serve as a model for the mechanistic behavior in cell crowds induced by
minorities, e.g., in leader cells \cite{Omelchenko2003,Yamaguchi2015}.

\section{Conclusions}

We modeled machines at nano- to micro-scales as particles with
an internal state and state-dependent shape.
Using the model, the interference between machines in a crowd
could be considered through the excluded volume.
Phase transition to a stalled state was observed to depend on the density
and fluctuations.
The reaction-diffusion behavior of such a machine crowd could be
investigated by explicitly introducing reactants as additional particles.
In combination with an enzymatic reaction model,
spatiotemporal patterns such as traveling waves were observed,
as was the segregation between active and inactive regions.
The system was proposed as a framework to model the molecular machinery
in a cell, but also as a model of crowds of cells or artificial machine-like objects.

\begin{acknowledgement}

The author thanks Hiroaki Takagi, Tatsuo Shibata, Masahiro Ueda, and Dan Tanaka for
their insightful comments at the early stage of this work.
This work was supported by MEXT and JSPS KAKENHI, grant numbers
JP20740243 and JP23115007 (``Spying minority in biological phenomena''),
and the Platform Project for Supporting Drug Discovery and Life Science Research
(Platform for Dynamic Approaches to Living System) from AMED, Japan.
Large-scale computer systems at the Cybermedia Center, Osaka University
were used in part for the simulations.
The author declares no competing financial interest.

\end{acknowledgement}

% \bibliography{achemso-demo}


%\begin{tocentry}
\begin{flushright}
\includegraphics[width=3.25in]{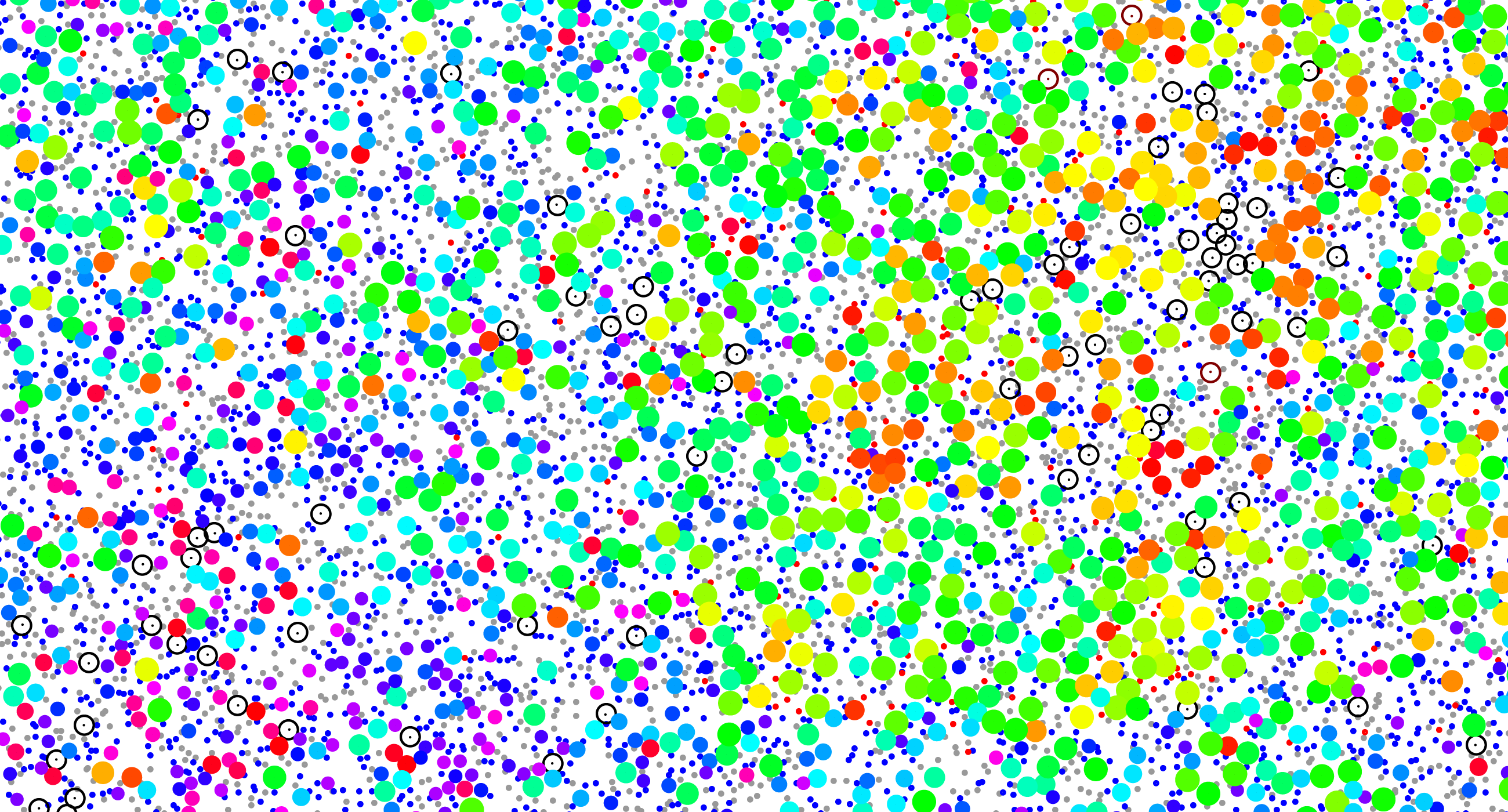}\\
\end{flushright}
%A snapshot of Model 2.
%\end{tocentry}

\end{document}